\documentclass[a4paper,12pt]{article}

\usepackage{a4wide}

\usepackage{graphicx}
\usepackage{dcolumn}
\usepackage{bbm}
\usepackage{epsfig}
\usepackage{amsmath}

\newcommand{\Tr}{\,\text{Tr}}
\newcommand{\I}{\text{i}}

\newcommand{\GQA}{\Gamma_{\text{QA}}}
\newcommand{\GQAr}{\Gamma_{\text{QA,R}}}
\newcommand{\WQA}{W_{\text{QA}}}
\newcommand{\Nf}{N_{\text{f}}}
\newcommand{\nL}{n_{\text{L}}}
\newcommand{\xI}{x_{\text{I}}}
\newcommand{\xF}{x_{\text{F}}}
\newcommand{\Sw}{S_{\text{w}}}
\newcommand{\vE}{v_{\text{E}}}

\title{\bf\large Quantum effective actions\\ from nonperturbative worldline
  dynamics}

\author{\large H. Gies \\ \\
\small \it Institute for theoretical physics, Heidelberg
  University,\\
\small \it Philosophenweg 16, D-69120 Heidelberg, Germany.\\ \\
\large J. S\'anchez--Guill\'en, R.A.~V\'azquez\\ \\
\small \it Instituto Galego de Altas Enerx\'{\i}as \\
\small \it Departamento de F\'{\i}sica de Part\'{\i}culas, \\
\small \it  Universidade de Santiago \\ 
\small \it 15782 Santiago de Compostela, Spain.}

\begin{document}

\maketitle

$\text{}$

\vspace{-14.3cm}

{\hfill \small\sf HD-THEP-05-12, 
{  }http://arXiv.org/abs/hep-th/0505275} 

\vspace{12.5cm}

\begin{abstract}
We demonstrate the feasibility of a nonperturbative analysis of
quantum field theory in the worldline formalism with the help of an
efficient numerical algorithm. In particular, we compute the effective
action for a super-renormalizable field theory with cubic scalar
interaction in four dimensions in quenched approximation (small-$\Nf$
expansion) to all orders in the coupling. We observe that
nonperturbative effects exert a strong influence on the infrared
behavior, rendering the massless limit well defined in contrast to the
perturbative expectation. Our numerical method is based on a direct
use of probability distributions for worldline ensembles, preserves
all Euclidean spacetime symmetries, and thus represents a new
nonperturbative tool for an investigation of continuum quantum field
theory.
\end{abstract}


\section{Introduction}

Many current problems in particle physics as well as statistical
physics demand for new nonperturbative field theoretic techniques.
Particularly, the treatment of strongly coupled fluctuations requires
methods that go beyond standard perturbative techniques.

Conventionally, perturbation theory is understood as an expansion for
both small coupling and small amplitude of the fields. The case
of larger fields but small coupling can be addressed with
effective-action techniques that take the coupling to all orders to an
external field into account; prominent examples are the
Heisenberg-Euler effective action \cite{Heisenberg:1935qt} or the
Coleman-Weinberg effective potential \cite{Coleman:1973jx}.

In the present work, we propose a technique that has the potential to
deal with the opposite limit: arbitrary coupling but weak
external-field amplitude\footnote{Even though the method takes
couplings to the external field into account to all-orders, the
approximations involved apply best to the case of weak fields, see
below.}. The technique is based on the worldline formulation of
quantum field theory that goes back to ideas of Feynman
\cite{feynman1}; it can be viewed as a mapping of field theoretic
problems onto quantum mechanical path integrals, identifying the
trajectory of a fluctuation in coordinate space with the path of a
quantum mechanical particle. The approach had occurred rarely in the
literature, see, e.g., \cite{Polyakov:ez}, until it was realized that
certain perturbative computations simplify tremendously in the
worldline approach \cite{berkos,strassler}; the deep reason behind
this observation is the fact, that the worldline formulation of field
theoretic correlators can be understood as the infinite string-tension
limit of string-theory amplitudes, giving rise to a higher level of
organization of the expressions. Important progress was made by
generalizing the worldline techniques to effective-action computations
\cite{Schmidt:1993rk,Schubert:2001he}. By now, the formalism has found
a wide range of application, see, for instance,
\cite{bedush,dunnor,cadhdu,Dilkes:1995cu,adlsch,Fosco:2003rr,gussho,Bastianelli:2004zp,marusc,Brummer:2004xc}.
The power of the worldline approach became particularly apparent in
combination with numerical Monte-Carlo techniques
\cite{Gies:2001zp,Schmidt:2002mt}; {\em worldline numerics} in the
form of efficient and fast algorithms is now available for the
computation of effective actions or Casimir energies for highly
general forms of the external field. For instance, nonlocal phenomena
and fluctuation-induced geometry-dependent interactions can now be
computed reliably and straightforwardly
\cite{Langfeld:2002vy,Gies:2003cv}.

The majority of applications of the worldline approach remains
perturbative in the coupling, even though formal all-order expressions
can already be found in the early works \cite{feynman1}. The present
work aims at exploiting these and related formal expressions for a
direct evaluation of all-order results by numerical Monte-Carlo
means. We demonstrate the potential of worldline numerics for
nonperturbative problems with the aid of a super-renormalizable
model field theory involving a ``charged'' real scalar field
$\phi$ coupled to a ``scalar photon'' $A$.\footnote{The attributes in
quotes serve only as an illustrative analogy, but should not be taken
literally, since the model does not have continuous local nor global
symmetries. Our model can also be viewed as a modified
Wick-Cutkosky \cite{WickCutkosky} model with imaginary coupling. It 
mimics as well the ubiquitous Yukawa coupling.}
Promoting the $\phi$ field formally to an $\Nf$-component field, we
study the system by means of an expansion for small $\Nf$. Already at
leading nontrivial order, this expansion corresponds to an infinite
set of Feynman diagrams involving one open $\phi$ line or,
alternatively, one closed $\phi$ loop but arbitrarily many scalar
photon fluctuations. We refer to the leading nontrivial order as the
``quenched approximation''.  This approximation holds for arbitrarily
large values of the coupling. Although it also includes infinitely
many couplings to the external field, we expect that the
approximation is most reliable for weak field amplitudes. As a first
concrete application, we compute the nonperturbative effective action for
the scalar photon, $\Gamma[A]$, in the quenched approximation,
concentrating on the induced interactions of soft photons. In this
way, we obtain an effective action of Heisenberg-Euler type to leading
order in $\Nf$ but to all orders in the coupling.

The main purpose of the present work is to demonstrate that our method
is capable of giving answers to nonperturbative problems with a
computation from first principles. With regard to more realistic
quantum field theories, our work serves as a feasibility study.
Nevertheless, even in the simple model we observe new phenomena that
may generalize to other theories as well. Of special relevance is the
following result: whereas the perturbative expansion is ill-defined
for a massless $\phi$ field (similar, e.g., to massless QED), the
nonperturbative effective action remains valid for massless fields,
revealing the unexpected breakdown of perturbation theory (and not of
the theory itself) in this limit.

Our work is related to other nonperturbative investigations of field
theoretic problems based on the worldline approach: in
\cite{Affleck:1981bm}, scalar QED (ScQED) in quenched approximation
was considered, and an instanton approximation of the path integral
was used to compute all-order corrections to the Schwinger
pair-production rate in a constant electric field. Furthermore, a
comprehensive study of the worldline expressions for various quenched
propagators has been performed in \cite{Rosenfelder:1995bd}, employing
a variational approach. Therein, the nonperturbatively interacting
path integrals are approximated by optimizing a trial action according
to a variational principle. In \cite{Sanchez-Guillen:2002rz}, the
worldline approach was used to deduce information about the
nonperturbative quenched propagator beyond the well-controlled
infrared (IR) limit. A combination of Monte-Carlo techniques with the
worldline approach has also been used in \cite{Savkli:1999rw} for the
determination of bound-state properties from 4-point correlators; see
also \cite{Brambilla:1997ky} for an approach to QCD bound states.
Moreover, non perturbative worldline methods have been used to calculate
the chiral anomalies \cite{Alvarez:1983},\cite{Fosco:2004}.

We start the presentation of our work in Sect.~\ref{NonpWM} by
rederiving the worldline expressions for the effective action and the
nonperturbative propagator in quenched approximation. Even though the
final results are known in the literature, we present their
derivation here (with details in App.~\ref{AppQA}) in the modern
language of generating functionals of correlation functions using
functional integrals. In Sect.~\ref{SecWorldlineN}, we elucidate the
numerical algorithm, concentrating on the new aspects introduced by
the present work, and presenting results for the free theory as a
useful example; another illustrative example is presented in App. C. In
Sect.~\ref{SecVpot}, we summarize our main
numerical results for the self-interaction potential of the worldlines
which lies at the heart of the nonperturbative dynamics.
Section \ref{SecGQA} employs this numerical data to derive our
main result, the effective action of the scalar photon. We explain how
the standard renormalization procedure applies to the present
nonperturbative study.
Section \ref{SecConc} summarizes our conclusions and gives an outlook.

\section{Nonperturbative worldline methods in QFT}
\label{NonpWM}

Let us consider a specific Euclidean
quantum field theoretic model in $D$ dimensional spacetime involving
two interacting real scalar fields $\phi$ and $A$ with bare Lagrangian
\begin{equation}
\mathcal L(\phi,A) =\frac{1}{2} (\partial_\mu \phi)^2 +\frac{1}{2} m^2 \phi^2
   +\frac{1}{2} (\partial_\mu A)^2 -\frac{\I}{2}\, h\, A\phi^2.
   \label{bareL}
\end{equation}
This model meets our requirements in many respects: (i) it has a
well-defined perturbative expansion, (ii) it will turn out to have
also a well-defined nonperturbative expansion in the quenched
approximation, (iii) the interaction $\sim h A\phi^2$ is
super-renormalizable in $D=4$, since the coupling has positive mass
dimension $[h]=1$, and hence facilitates a simple analysis of the
(nonperturbative) divergencies, (iv) the imaginary interaction imitates
the interaction occurring in QED with $A$ being the scalar analogue of
the massless photon field; hence, $A$ will be called the ``scalar
photon'' in the following. We are well aware of the fact that this
Lagrangian does not define a fully consistent QFT, since the potential
is not bounded from below; however, contrary to the corresponding
model with a real interaction (the Wick-Cutkosky model), the present
model does not have propagator singularities which would signal
the metastability of the model immediately and require a
careful treatment. Nevertheless, the Euclidean effective
action will turn out to exhibit nonzero imaginary parts which suggest
a relation to decay widths of the metastable states. Note that in
\cite{Bender:1999ek} it was argued that, despite the obvious
non-hermiticity of the Hamiltonian of this model, the spectrum could
be positive definite by virtue of a careful implementation of PT
symmetry.

\begin{figure}[hbt]
\centering
\mbox{\epsfig{figure=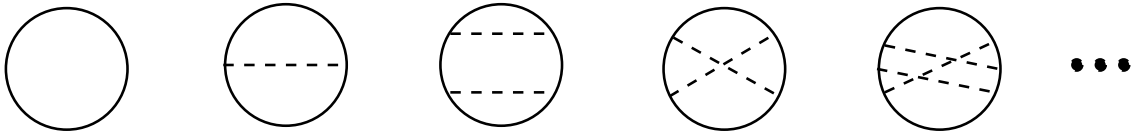,width=15.0cm}}
\caption{Diagrams included in the quenched approximation.}
\label{figQA}
\end{figure}

Let us first concentrate on the effective action $\Gamma[A]$ for soft
scalar photons with momenta $p^2$ smaller than the scale of the
massive $\phi$ field, $p^2\ll m^2$. This effective action can be
derived from the Schwinger functional,
\begin{eqnarray}
Z[J]\equiv e^{W[J]} 
&=&\int \mathcal D \hat A \mathcal D \phi \, e^{-\int_x \mathcal
  L(\phi,\hat A) + \int_x J \hat A} \nonumber\\
&=&\int \mathcal D \hat A\, \det{}^{-1/2} \big( -\partial^2 + m^2 +\I h
  \hat A\big)\, e^{-\int_x \frac{1}{2} (\partial_\mu \hat A)^2 + \int_x J
  \hat A} \nonumber\\
&=& \int \mathcal D \hat A \, e^{-S_{\text{eff}}(\hat A) +
  \int_x J \hat A},
\label{SchwingerF}
\end{eqnarray}
where $\int_x\equiv \int d^D x$ and $S_{\text{eff}}$ denotes an
auxiliary effective action governing the quantum dynamics of the
scalar photon (it should not be confused with the full quantum
effective action $\Gamma[A]$ defined below). Here, we have integrated
out the massive $\phi$ field, resulting in a functional determinant that
can be displayed in worldline form:
\begin{eqnarray}
S_{\text{eff}}(\hat A)&=& \int_x \frac{1}{2} (\partial_\mu \hat A)^2 +
\frac{1}{2} \ln \det  \big( -\partial^2 + m^2 +\I h
  \hat A\big) \nonumber\\
&=&\int_x \frac{1}{2} (\partial_\mu \hat A)^2 -
\frac{1}{2}\int_0^\infty \frac{dT}{T} \, e^{-m^2 T}
\, \Tr\, e^{( -\partial^2 +\I h  \hat A )} \nonumber\\
&=& \int_x \frac{1}{2} (\partial_\mu \hat A)^2 -
  \frac{1}{2(4\pi)^{D/2}}\int_0^\infty \frac{dT}{T^{1+D/2}} \,
  e^{-m^2  T}
  \Big\langle e^{\I h \int_0^T d\tau \, \hat A(x(\tau))} \Big\rangle_x.
\label{Seff}
\end{eqnarray}
Here, the expression $\langle\dots\rangle_x$ denotes an expectation value
with respect to an ensemble of worldlines $x(\tau)$ with a Gau\ss ian
velocity distribution,
\begin{equation}
\langle \dots \rangle_x
  = \frac{ \int \mathcal D x \,\dots\,  e^{-\frac{1}{4} \int_0^T d\tau\,
           \dot x^2} }
  { \int \mathcal D x \,  e^{-\frac{1}{4} \int_0^T d\tau\,
           \dot x^2} }.
\label{Gaussv}
\end{equation}
In a standard fashion, the effective action $\Gamma[A]$ -- the
generating functional of all 1PI correlation functions of the scalar
photons -- is related to the Schwinger functional by a Legendre
transformation,
\begin{equation}
\Gamma[A]=- W[J] + \int_x J A, \label{Legendre}
\end{equation}
with $J=J[A]$ being a functional of the ``classical'' field $A$
(conjugate to the source $J$) defined by
\begin{equation}
A(x):= \frac{\delta W[J]}{\delta J(x)}\equiv A[J]. \label{classA}
\end{equation}
In this work, we will evaluate $\Gamma[A]$ in the so-called quenched
approximation. In a diagrammatic language, the quenched approximation
for $\Gamma[A]$ can be characterized by taking into account all
diagrams with exactly one closed $\phi$ loop, but arbitrarily many
virtual scalar photon fluctuations, see Fig.~\ref{figQA}. From a
different viewpoint, the quenched approximation would correspond to
the lowest nontrivial order in an $\Nf$ expansion if we promoted the
$\phi$ field to have $\Nf$ components.

More formally, let us introduce the $\phi$ field worldline current
\begin{equation}
j(x):=\I h \int d\tau\, \delta^{(D)}(x-x(\tau)), \label{jcurr}
\end{equation}
such that the interaction of the scalar photon $A$ with the $\phi$
worldline  can be expressed as
\begin{equation}
\Big\langle e^{\I h \int_0^T d\tau \, A(x(\tau))} \Big\rangle_x
=\Big\langle e^{\int_x j A} \Big\rangle_x. \label{Ajcurr}
\end{equation}
Then the quenched approximation consists in dropping all terms
involving worldline correlators between two different $j$ currents of
the $\phi$ field, $\mathcal O ( \langle j_1 \dots j_2
\rangle_{x_1,x_2}) \to 0$. Since each current couples to all orders to
the classical field $A$, we expect this approximation to be good if
the scalar photon field under consideration is weak compared to other
mass scales of the system. Nevertheless, the quenched approximation
does not restrict the value of the coupling $h$ which therefore can be
arbitrarily large.

In appendix \ref{AppQA}, we derive the general worldline
representation of the Schwinger functional and perform the Legendre
transform yielding the effective action. In the quenched approximation,
$\Gamma[A]$ can be displayed in closed form (cf. Eq.~\eqref{derGQA}),
\begin{equation}
\GQA[A]= \int_x \frac{1}{2} (\partial_\mu A)^2
-\frac{1}{2(4\pi)^{D/2}} \int_0^T \frac{ dT}{T^{1+D/2}}\,
e^{-m^2T}
\Big\langle e^{\frac{1}{2}\int_x j \Delta j +\int_x j A} \Big\rangle_x,
\label{GQA1}
\end{equation}
where $\Delta=1/(-\partial^2)$ denotes the propagator of the scalar
photon with coordinate space representation
\begin{equation}
\Delta(x_1,x_2) = \frac{ \Gamma \left(\frac{D-2}{2} \right)}
      {4\pi^{D/2}} \, \frac{1}{|x_1-x_2|^{D-2}}.  \label{Delta}
\end{equation}
It is important to note that, in particular, all contributions from
the scalar photon fluctuations are summarized in the exponential with a
worldline current-current interaction $\sim \int_x j\Delta
j$. Re-expressed in terms of the worldline coordinates $x(\tau)$ of the
$\phi$ field fluctuations, this term acts like a self-interaction
potential $V[x]$ of the worldline,
\begin{eqnarray}
\frac{1}{2} \int_x j \Delta j &=& - \frac{h^2}{8 \pi^{D/2}} \,
\Gamma\left({\scriptstyle \frac{D-2}{2}} \right) \int_0^T d\tau_1
d\tau_2\, \frac{1}{|x_1 -x_2|^{D-2}} \nonumber\\
&=:& - g \, V[x], \quad \text{with}\,\,g:=\frac{h^2}{8 \pi^{D/2}} \,
\Gamma\left({\scriptstyle \frac{D-2}{2}} \right). \label{Vpot}
\end{eqnarray}
Here, we have introduced an effective coupling $g$ for convenience.
Therefore, a thorough analysis of the properties of this
potential and its vacuum expectation value is mandatory for an
understanding of the nonperturbative dynamics. This is one of the main
aims of the present work.
In addition to the effective action for the scalar photon, also proper
vertices with external $\phi$ lines can be studied nonperturbatively
on the worldline. In particular,  the computation of the propagator
of the $\phi$ field is straightforward 
and its worldline representation can be compactly written as
\begin{equation}
G(\xF,\xI)=\frac{1}{(4\pi)^{D/2}} \int_0^\infty \frac{dT}{T^{D/2}}\,
e^{-m^2T}\, e^{-\frac{|\xF-\xI|^2}{4T}}\, \Big\langle e^{-g V[x]}
\Big\rangle_{\xI}^{\xF}. \label{Gfullx2}
\end{equation}
Here, the expectation value $\langle\dots\rangle_{\xI}^{\xF}$ implies
that we are dealing with open worldlines ranging from $\xI$ to
$\xF$. Again a proper understanding of the properties of the potential
term is required.

Furthermore, we would like to point out that the $\phi$ propagator can
even be generalized to the case of arbitrarily many couplings to an
external scalar photon background $A$ by including a factor of $e^{\I
  g\int_0^T d\tau A(x(\tau))}$ in the expectation value of
Eq.~\eqref{Gfullx2}.

Let us conclude this section by translating the present results to
scalar quantum electrodynamics (ScQED). In this case, the bare
Lagrangian is given by
\begin{equation}
\mathcal L_{\text{ScQED}}= | D_\mu[A] \phi |^2 + m^2 |\phi|^2 +
\frac{1}{4} F_{\mu\nu} F_{\mu\nu}, \label{LScQED}
\end{equation}
where $\phi$ now denotes a complex scalar field and $D_\mu[A]=
\partial_\mu +\I e A_\mu$ is the covariant derivative. Most of our
formulas can immediately be generalized to this case with the
effective photon action given by
\begin{equation}
\GQA[A_\mu]= \int_x \frac{1}{4} F_{\mu\nu}F_{\mu\nu} -
\frac{1}{(4\pi)^{D/2}} \int_0^\infty \frac{dT}{T^{1+D/2}}
  \Big\langle e^{\frac{1}{2} \int_x j_\mu \Delta_{\mu\nu} j_\nu +
    \int_x j_\mu A_\mu} \Big\rangle_x,
\label{GQAScQED}
\end{equation}
with the worldline current of the charged field
\begin{equation}
j_\mu(x)=\I e \int_0^T d\tau\, \dot x_\mu(\tau) \,
\delta^{(D)}(x-x(\tau)), \label{jcurrScQED}
\end{equation}
and the photon propagator
\begin{eqnarray}
\Delta_{\mu\nu}(x_1,x_2)&=& \frac{\Gamma(\frac{D-2}{2})}{4\pi^{D/2}}
\left[ \frac{1+\alpha}{2} \frac{1}{|x_1-x_2|^{D-2}}
+({\scriptstyle \frac{D}{2}}-1)(1-\alpha) \frac{(x_1-x_2)_\mu
  (x_1-x_2)_\nu}{|x_1-x_2|^D} \right],
\label{photprop}
\end{eqnarray}
with gauge parameter $\alpha$. The interaction between the charged
worldline current and the photon field then occurs in the form of a
Wegner-Wilson loop,
\begin{equation}
e^{\int_x j_\mu A_\mu}=e^{\I e \oint dx_\mu A_\mu(x)}. \label{WWloop}
\end{equation}
Here, the current-current interaction induced by the virtual photon
exchanges inside the $\phi$ loop now implies a worldline
self-interaction potential of the form
\begin{equation}
\frac{1}{2} \int_x j_\mu \Delta_{\mu\nu} j_\nu
= -\frac{e^2}{8 \pi^{D/2}} \Gamma( {\scriptstyle \frac{D-2}{2}})
V_{\text{ScQED}}[x]
\label{ccintScQED}
\end{equation}
with
\begin{equation}
V_{\text{ScQED}}[x]= \int_0^T d\tau_1 d\tau_2\, \left[
  \frac{1+\alpha}{2} \frac{\dot x_1 \cdot \dot x_2}{|x_1-x_2|^{D-2}}
  + ({\scriptstyle \frac{D}{2}}-1)(1-\alpha) \frac{\dot x_1\cdot (x_1-x_2)
  (x_1-x_2)\cdot\dot x_2}{|x_1-x_2|^D} \right].
\label{VpotScQED}
\end{equation}
Note that the second term drops out completely in the Feynman gauge
with $\alpha=1$. In fact, all $\alpha$-dependent terms can be shown to
correspond to a total derivative that automatically vanishes in the
case of the photon effective action where all worldlines are closed loops
\cite{Schubert:2001he}.

The representation of the scalar propagator in Eq.~\eqref{Gfullx2}
does also hold in the case of ScQED with $V[x]$ replaced by
$V_{\text{ScQED}}[x]$ of Eq.~\eqref{VpotScQED} and
$g_{\text{ScQED}}=\frac{e^2}{8 \pi^{D/2}} \Gamma( {\scriptstyle
  \frac{D-2}{2}})$.

In the case of ordinary spinor QED, an additional complication arises
from the coupling between fermionic spin and the electromagnetic
field. Using a convenient spin-factor representation, the present
formulation can be shown to hold with the simple insertion of the spin
factor in all worldline expectation values, see
\cite{GiesHaemmerling}, \cite{Fosco:prep}.

\section{Worldline numerics}
\label{SecWorldlineN}

In this section, we develop the discretization of the integral in the
proper time. Since the discretization is done for an auxiliary
variable, i.e., an integration parameter, it has the advantage of
preserving the continuum spacetime symmetries, i.e., the Lorentz
(Euclidean) invariance. Also gauge and chiral symmetries would be
preserved if present. This is the major difference between
nonperturbative worldline numerics and other Monte-Carlo methods such
as lattice field theory. In this sense, our method is much closer to
analytic continuum field-theory techniques than to numerical lattice
formulations.

\subsection{Algorithmic principles}
\label{algorithmic}

In this work, we evaluate the worldline integrals, corresponding to
quantum mechanical path integrals, by means of Monte-Carlo algorithms,
as proposed for perturbative amplitudes in \cite{Gies:2001zp}. For
this, we approximate the infinite set of worldlines by a finite
ensemble with $\nL$ configurations. Furthermore, we
discretize the propertime parameter of each worldline, such that the
fluctuating worldline is specified by a set of $N$ points per line
(ppl),
\begin{eqnarray}
\langle \dots \rangle_x &=&\frac{\int \mathcal D x \dots P[x]}
{\int \mathcal D x P[x]} \, \to \frac{1}{\nL} \sum_{\{ x\}}
\dots, 
\quad\text{with}\quad P[x]= e^{-\frac{1}{4} \int_0^T d\tau \dot{x}^2},
\nonumber\\
x(\tau) &\to& x_i \in \mathbbm{R}^D,\quad i=0,1,\dots,N.\label{WN}
\end{eqnarray}
In the case of the effective action, the worldlines form closed loops
and we identify $x_0$ with $x_N$; in the case of the $\phi$
propagator, where the initial and final point of the worldline are
fixed, we use the definition $x_0=\xI$ and $x_{N+1}=\xF$.

The finite number of worldlines $\nL$ introduces a statistical error
that will be estimated automatically by the Monte-Carlo algorithm. The
finite number of points per worldline $N$ implies a systematic error
that has to be analyzed straightforwardly by approaching the
propertime continuum limit $N\to \infty$. This is the most sumptuous
point of the present investigation.

An immediate realization of Eq.~\eqref{WN} requires to generate a
different ensemble for each value of the propertime parameter $T$,
occurring in the Gau\ss ian velocity distribution. This can elegantly
be circumvented by introducing {\em unit loops} or {\em lines} $y(t)$
\cite{Gies:2001zp},
\begin{equation}
y(t):=\frac{1}{\sqrt{T}} \, x(Tt), \quad t\in [0,1],
\label{unitl1}
\end{equation}
such that the velocity distribution for $y(t)$ becomes independent of
$T$,
\begin{equation}
\int_0^T d\tau\, \dot x^2(\tau) = \int_0^1 dt\, \dot y^2(t).
\label{unitl2}
\end{equation}
Here, the dot always denotes a derivative with respect to the
corresponding argument. As a consequence, one and the same ensemble of
unit loops or lines can be used for all values of $T$, saving an
enormous amount of CPU time. Of course, this change of variables also
affects the remaining worldline integrand, for instance, the
self-interaction potential,
\begin{eqnarray}
V[x]&=&\int_0^T d\tau_1 d\tau_2\, \frac{1}{|x_1-x_2|^{D-2}}
=T^{3-D/2}\, \int_0^1 dt_1 dt_2\, \frac{1}{|y_1-y_2|^{D-2}},
\nonumber\\
&=:& T^{3-D/2}\, v[y]\label{vpot}
\end{eqnarray}
where we have introduced the dimensionless self-interaction potential
of the unit loops, $v[y]$.\footnote{Incidentally, we note that the same
transformation implies for the potential in ScQED that
$V_{\text{ScQED}}[x] =T^{2-D/2} v_{\text{ScQED}}[y]$; therefore,
$V_{\text{ScQED}}$ is already dimensionless in $D=4$ and the unit-loop
transformed exponential of the potential becomes independent of $T$.}
For computations of the $\phi$ propagator, the variable transformation
also affects the boundary points, inducing a further $T$ dependence,
$\xF=\sqrt{T}y_{\text{F}}$, $\xI=\sqrt{T}y_{\text{I}}$, that can
nevertheless be handled straightforwardly.

In the context of the effective-action calculation involving closed
worldlines, one final convenient step for the algorithm consists in
using the center-of-mass decomposition of the worldlines,
$x(Tt)=x_{\text{CM}}+ \sqrt{T} y(t)$, such that the unit loops satisfy
$\int_0^1 dt\, y(t)=0$. This implies that the ensemble average becomes
\begin{equation}
\langle \mathcal F[x] \rangle_x= \int d^D x_{\text{CM}} \, \langle
\mathcal F[x_{\text{CM}}+\sqrt{T} y] \rangle_y. \label{cmdecomp}
\end{equation}
This allows to study the effective action density, i.e., the effective
Lagrangian, locally, with the center-of-mass integration corresponding
to the usual relation between action and Lagrangian,
$\Gamma=\int d^D x_{\text{CM}} \mathcal L$.

For the Monte-Carlo generation of the worldline ensembles, there are
various efficient techniques available that do not require to spend
computer time on dummy thermalization sweeps, as it would be the case
for standard heat-bath algorithms. Fast algorithms are, for instance,
the Fourier loop algorithm, ``f loops'', or explicit diagonalization
of the velocity distribution, ``v loops'', see \cite{Gies:2003cv}. For most
computations in the present work, we use a new algorithm that is based
on a point doubling procedure (``d loops'' and ``d lines'') that is
described in appendix \ref{Appdloops}.

Beyond these general algorithmic aspects, we encounter a particular
numerical problem in the present case upon discretizing the
self-interaction potential \eqref{vpot}: the integrand for coinciding
points $y_1=y_2$ is ill-defined. The possibly arising divergence can
be viewed as a ``self-energy'' of the worldline. We regularize the
discretized sum by taking out the coincident points by hand,
\begin{equation}
v[y] \to \frac{1}{N^2} \sum_{i\neq j} \frac{1}{|y_i -y_j|^{D-2}}, \quad,
i,j=0,\dots N. 
\label{vreg}
\end{equation}
A singular behavior of the original integral then manifests itself in
a strong $N$ dependence of $v[y]$ which we will carefully
control. Since the distance of the nearest-neighbor points decreases
for increasing $N$, $|y_i-y_{i+1}| \sim 1/\sqrt{N}$, such strong $N$
dependence signals a short-distance UV singularity that will be
subject to renormalization of physical parameters. Alternatively, we
could include the coincident points in the above sum by shifting the
denominator, $|y_i-y_j| \to |y_i-y_j| + \epsilon$, and study the
limiting behavior with $\epsilon \to 0$. However, since the latter
limit will always interfere numerically with the large-$N$ limit
(which has to be taken anyway), the first proposed method can be
handled more straightforwardly.

\subsection{Free theory}

As a first useful application, let us study the worldline representation
for the
free theory in the limit of vanishing interaction, e.g., $h\to0$ for
the purely scalar system.
An instructive example is given by the kinetic action for the
worldlines, $\Sw$, characterizing the Gau\ss ian probability
distribution of the worldline ensemble,
\begin{equation}
P[x]= \bar{\cal N} \exp(-\Sw[x] ),
\end{equation}
where $\bar{\cal N}$ is the normalization, ensuring $\int \mathcal D
x\, P[x]$=1. Discretizing the free worldline action $\Sw$ (introduced
in Eq.~\eqref{Gaussv}), as outlined above for the case of the $\phi$
propagator, we obtain
\begin{equation}
\Sw=\frac{1}{4} \int_0^T d\tau \dot x^2(\tau) \to
 \frac{N+1}{4T}  \sum_{i=1}^{N+1} (x_i-x_{i-1})^2,
\label{Sw}
\end{equation}
with $x_0=\xI$ and $x_{N+1}=\xF$ the corresponding initial and final
points of the $\phi$ propagation. For the propagator, we would finally
have to integrate over the propertime $T$. 
However, we will be satisfied here with the
propertime integrand and set $T$ to some fixed value in the following.

The ensemble average and the value of the root mean square (RMS) of
the free worldline action can be calculated analytically.
For this, let us consider the auxiliary integral
\begin{equation}
I_N(\xF,\xI,T,\kappa) = \int dx_1 \cdots dx_N \exp\{- \frac{\kappa
  \Delta \tau }{4} \sum_{i=1}^{N+1} \frac{(x_{i}-x_{i-1})^2}{\Delta
  \tau^2} \},
\end{equation}
where $\Delta \tau= T/(N+1)$, and we have suppressed the Lorentz indices
of the $x_i$'s which are points in $D$ dimensional spacetime. The
parameter $\kappa$ is arbitrary here which will be exploited below.

The integral is Gau\ss ian and can be done exactly. We rewrite the
integral as
\begin{equation}
I_N(\xF,\xI,T,\kappa) = \int dx_1 \cdots dx_N
  \exp\{- \frac{\kappa \Delta \tau }{4}
\frac{1}{\Delta \tau^2} ( \xI^2 + \xF^2 - 2 \xI x_1 -2 \xF x_N
  + x_i A_{ij} x_j ) \},
\end{equation}
where $A$ is a symmetric matrix,
\begin{equation}
A_{ij} = 2 \delta_{ij} - \delta_{i+1,j} - \delta_{i,j+1}, \quad
i,j=1,\dots, N.
\end{equation}
Introducing the auxiliary current $\eta_i$,
\begin{equation}
\eta_i = -2 \xI \delta_{i1} -2 \xF \delta_{iN},
\end{equation}
the integral reads
\begin{eqnarray}
I_N(\xF,\xI,T,\kappa)
=  (\frac{4 \pi \Delta \tau}{\kappa})^{\frac{N D}{2}}
\frac{1}{\sqrt{\det(A)}^D } \exp\left[ \frac{\kappa}{4 \Delta \tau }
\frac{\eta_i (A^{-1})_{ij} \eta_j}{4}  
-\frac{\kappa}{4 \Delta \tau} (\xI^2+ \xF^2) \right],
\label{aveaction}
\end{eqnarray}
with
\begin{equation}
\eta_i (A^{-1})_{ij} \eta_j = 4 (\xI^2 (A^{-1})_{11} + 2 \xI \xF
(A^{-1})_{1N} + A^{-1}_{NN} \xF^2 ).
\end{equation}
The ensemble average of the free action can now be calculated from the
auxiliary integral,
\begin{equation}
\langle \Sw\rangle = \bar{\cal N} \int dx_1 \cdots dx_N \; \Sw \;
  \exp\{ - \Sw\} =
\bar{\cal N} \int dx_1 \cdots dx_N \; \frac{-d}{d\kappa} \; \exp\{
  -\kappa S_w\}\bigg|_{\kappa=1}.
\end{equation}
The integral actually agrees with $I_N$ of Eq.~\eqref{aveaction} above,
therefore
\begin{eqnarray}
\langle\Sw\rangle &=& \bar{\cal N} \; \frac{-d}{d\kappa}
 \left(\frac{4 \pi \Delta \tau}{\kappa}\right)^{\frac{N
D}{2}} \frac{1}{(\sqrt{\det A})^D} \exp\{ - \frac{\kappa}{4 T}
 (\xF-\xI)^2\}\bigg|_{\kappa=1} \nonumber\\
&=& \frac{1}{2} N \, D + \frac{1}{4T} (\xF -\xI)^2.\label{SwVEV}
\end{eqnarray}
Notice that the result corresponds to the classical action plus one
half times the number of degrees of freedom.

\begin{figure}[hbt]
\centering
\mbox{\epsfig{figure=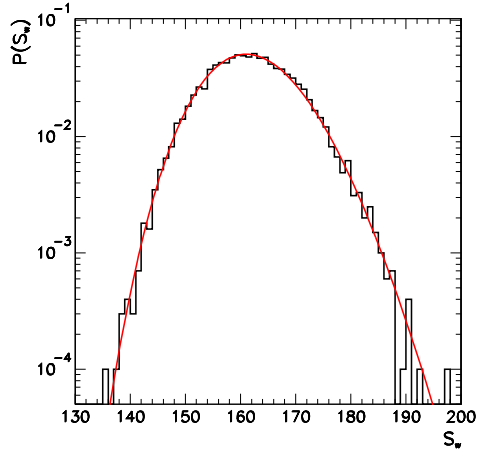,width=7.0cm}}
\caption{Action distribution from the Monte-Carlo method (histogram)
  in comparison with the analytical calculation (continuous line) for
  $N=31$ and $\xF=(10,0,10,0)$.}
\label{fig:actiondis}
\end{figure}

In table \ref{table:aveaction}, we show a comparison between the exact
results and those of the numerical calculation using our Monte-Carlo
code for different parameters. The numerical result is always very
satisfactory with a precision better than 0.1 \% in all cases.
This illustrates the power of the Monte-Carlo method.
\begin{table}
\begin{center}
\begin{tabular}{|cccccc|}
\hline
$N$  & $2T$ & $\xF$ & $\langle\Sw\rangle_{\rm numerical}$ &
$\langle\Sw\rangle_{\rm theory}$ &
$\Delta \Sw/\Sw (\%)$ \\
\hline
31   & 0.5 & (10,0,10,0) & 161.895 & 162   & -0.06  \\
1023 & 0.5 & (0,0,0,0)   & 2046.38 & 2046  & 0.02   \\
1023 & 0.5 & (10,0,0,0)  & 2096.00 & 2096  & 0.00   \\
4095 & 0.5 & (0,0,0,0)   & 8190.15 & 8190  & 0.001  \\
4095 & 0.5 & (10,0,0,0)  & 8238.70 & 8240  & -0.02  \\
4095 & 0.5 & (100,0,0,0) & 13189.4 & 13190 & -0.005 \\
\hline
\end{tabular}
\end{center}
\caption{Average theoretical and numerical actions. All numerical
  calculations were done with $\nL=10^4$ worldlines. We have set 
$D=4$ and $\xI=(0,0,0,0)$.}
\label{table:aveaction}
\end{table}
We underline a fact that will be of the outmost importance in the more
involved non perturbative calculation below, namely that
the Monte-Carlo method also provides information about the probability
distribution of the free action itself. For instance, the probability
of finding a configuration with worldline action  $\Sw'$ is given by
\begin{equation}
{\cal P}(\Sw') = \bar{\mathcal{N}} \int dx_1 \cdots dx_N  \,
\delta(\Sw-\Sw') \, \exp\{ - \Sw\},
\end{equation}
which can be evaluated analytically, yielding
\begin{equation}
{\cal P}(\Sw') = \bar{\cal N} (\Sw'-S_{\rm class})^{-1+\frac{N D}{2}}
e^{-(\Sw'- S_{\rm class})} \; \theta(\Sw'-S_{\rm class}),
\end{equation}
where $S_{\rm class}= (1/4T) (\xF-\xI)^2$ is the classical action. In
Fig. \ref{fig:actiondis}, we compare the action distribution
calculated with our Monte-Carlo code to the analytical result.  The
agreement with the theoretical calculation is excellent.\footnote{We
  have chosen $N$ small to exhibit more clearly the non Gau\ss ian
character of the distribution}
Using the probability distribution ${\cal P}(\Sw')$, it is
straightforward to calculate the ensemble average, the RMS, and higher
moments of the free action. The ensemble average immediately verifies
Eq.~\eqref{SwVEV}, whereas the RMS results in
\begin{equation}
\sigma^2(\Sw) = \langle\Sw^2\rangle-\langle\Sw\rangle^2 = \frac{N D}{2}.
\end{equation}
As $N$ increases, the ensemble average $\Sw(\xF,\xI)$ grows as well,
and the distribution of $\Sw$ broadens. For large $N$, we observe from
\begin{equation}
\frac{\sigma(\Sw)}{\langle \Sw\rangle} \rightarrow \frac{1}{\sqrt{N D/2}}
\end{equation}
that the relative width of the distribution becomes narrower.

\section{Self-interaction potential in the scalar model}

\label{SecVpot}

In this section, we perform a detailed numerical analysis of the
self-interaction potential, occurring in Eq.~(\ref{Vpot}). In
particular, we consider its dimensionless form $v[y]$ defined in
Eq.~\eqref{vpot} in its discretized regularized version, 
\begin{equation}
v[y] = \frac{1}{N^2} \sum_{i\neq j} \frac{1}{|y_i-y_j|^{D-2}},
\end{equation}
as discussed in Eq.~\eqref{vreg}.  In the following, we
concentrate on $D=4$ dimensional spacetime. Other dimensions can be
treated in the same way.

\begin{figure}[hbt]
\centering
\mbox{\epsfig{figure=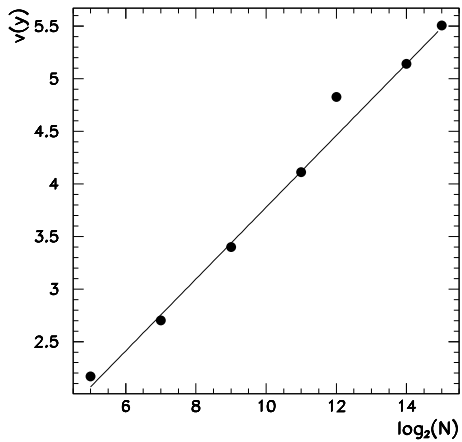,angle=0,width=7.0cm}}
\caption{Average value of the self-interaction potential as a
  function of the logarithmic of the number of points per loop ($N$).}
\label{fig:svev}
\end{figure}

\begin{figure}[hbt]
\centering
\mbox{\epsfig{figure=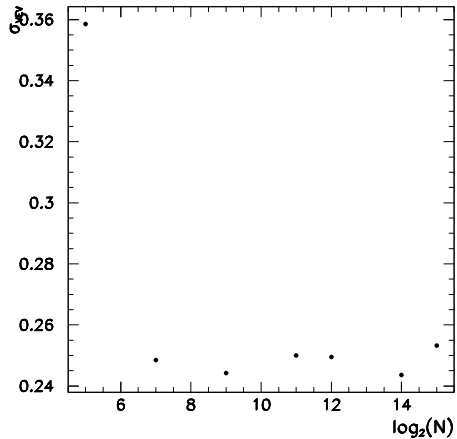,angle=0,width=7.0cm}}
\caption{RMS  of the scalar to scalar potential as a
  function of the number of points per loop ($N$).}
\label{fig:sigmasvev}
\end{figure}

The average value of the self-interaction potential $v[y]$ as a function of
the number of points per loop can be seen in
Fig.~\ref{fig:svev} together with a logarithmic fit. Obviously,  the
self-interaction potential diverges logarithmically with $N$,  
\begin{equation}
\langle v\rangle_y = a + b \ln(N), \label{vvev1}
\end{equation}
with $a \simeq 0.363 $ and $b\simeq 0.341$. By contrast, the RMS is finite, as
is visible in Fig.~\ref{fig:sigmasvev}. We have furthermore
collected numerical evidence that all the higher-order moments are
also finite. 

\begin{figure}[hbt]
\centering
\mbox{\epsfig{figure=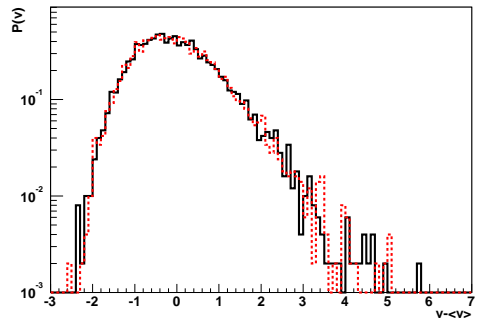,angle=0,width=7.0cm}}
\caption{Distribution of potentials for $N= 2^{9}$ (continuous line)
and $N=2^{12}$ (dashed line).
The average values of the distributions have been shifted to lie on
top of each other.}
\label{fig:spotdist_uv}
\end{figure}

In order to evaluate
ensemble averages, it will become useful to study the distribution of the
self-interaction potential defined by
\begin{equation}
P(v) = \hat{\cal N} \int {\cal D}y \; \delta(v-v[y]) \;
\exp(-\Sw[y]-v[y]), \label{Vdist}
\end{equation}
where the normalization is fixed such that $\int dv\, P(v)=1$. We
depict the numerical result for this distribution in
Fig.~\ref{fig:spotdist_uv} with binned values of the potential for
two worldline ensembles with different $N$ ppl. The
distributions are shifted along the $x$ axis by the logarithmically
divergent ensemble average of $v$, but are not modified otherwise. As a
result, all the distributions are equal within the numerical accuracy
for different values of $N$, demonstrating that all other
properties of the distribution are devoid of further divergencies.

This remarkable property reflects the fact that we are dealing with a
super-renormalizable interaction. In a diagrammatic language, only a
finite subset of diagrams is superficially divergent. In the present
scalar model, this finite subset consists of the one-loop contribution
to the $\phi$ 2-point function (mass operator), the one- and two-loop
scalar photon tadpole and the photon 2-point function (``vacuum
polarization''). Once these diagrams are controlled by adjusting the
counterterms and renormalizing the corresponding physical parameters,
all higher-loop diagrams do not contribute independently to further
renormalization; divergencies in subdiagrams will be canceled by the
already adjusted counterterms. Evidently, our nonperturbative
evaluation of the infinite set of diagrams contributing to the
quenched approximation will also benefit from the simplified
renormalization in this particular scalar theory.

Let us show that all the characteristics of the average potential can
be understood in terms of the distribution of chord distances and that
this distribution gives an intuitive picture of the problem.
\begin{figure}[hbt]
\centering
\mbox{\epsfig{figure=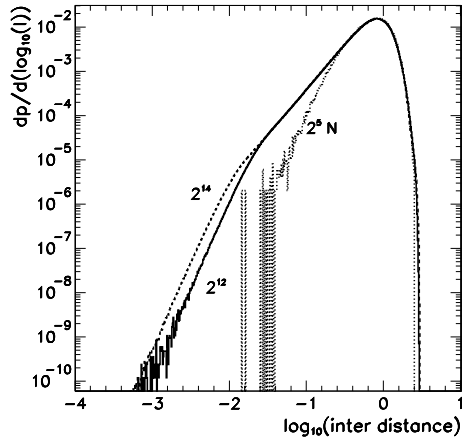,angle=0,width=7.0cm}}
\caption{Chord-distance distribution between pair of points in paths
for several values of points per loop, as marked.}
\label{fig:interdis_uv}
\end{figure}
For a large number of points per loop we can introduce the chord-distance
distribution, 
\begin{equation}
v[y] = \frac{1}{N^2} \sum_{i\neq j} \frac{1}{l_{ij}^{2}} = \int dl \, p(l) 
\frac{1}{l^2},
\end{equation}
where $l_{ij}=|y_i-y_j|$ abbreviates the chord distances and $p(l)$ is the
probability of having two points separated by a distance $l$.

In Fig.~\ref{fig:interdis_uv}, we show the chord-distance
distribution in a double-log plot for $N=2^5, 2^{12}$, and $2^{14}$.
From the figure, we can see that very large chord distances are
exponentially suppressed. For very small values of the distance, $l
\ll 1/\sqrt N$, much less than a typical random step, we see that the
probability goes like $l^{D-1}$ which is simply a measure of the
available volume in $D$ dimensions. This behavior is a direct
consequence of the worldline discretization, and the corresponding
branch vanishes in the continuum limit. For intermediate distances the
distribution goes like $l^1$ which implies that the path spans a 2
dimensional surface, characteristic of a random walk.  The
contribution to the average potential of the very low chord-distance
region is finite, in 4 dimensions.  However the intermediate region
gives a contribution of order
\begin{equation}
\int_{1/\sqrt N} dl\, p(l) \frac{1}{l^2} \sim -\ln(1/\sqrt N),
\end{equation}
which diverges in the continuum limit and gives rise to divergencies
in the average potential and, consequently, in the effective action.
The divergent part of the average potential carries information about
the topological properties of the paths which contribute to the
average.

\section{Effective action for the scalar photon}
\label{SecGQA}

Let us now exploit our knowledge about the worldline self-interaction
potential in the scalar model in order to compute the nonperturbative
effective action $\Gamma[A]$ of the scalar photon in quenched
approximation in $D=4$ spacetime dimensions.  In this section, we
demonstrate how the renormalization procedure can be carried out to
all orders, using this example.  For this, we need to consider the
properties of the potential $V[x]$ for closed worldlines, or its
dimensionless counterpart for unit loops, $v[y]$,
cf. Eq.~\eqref{vpot}.

In addition to the quenched
approximation, we confine ourselves to slowly varying photon fields
with frequencies smaller than the $\phi$ mass scale, $\omega^2\ll
m^2$, for which $A$ can be assumed constant in spacetime. Compared to
QED, these assumptions are reminiscent to those leading to the famous
Heisenberg-Euler effective action. Notice that this approximation is
consistent with the quenched approximation.

Employing the representation \eqref{GQA1} together with
Eq.~\eqref{Vpot} and the unit-loop transformation of
Eqs.~\eqref{unitl1}-\eqref{vpot}, the effective action in these limits
reads in $D=4$,
\begin{equation}
\GQA[A]= -\frac{1}{32\pi^2} \int_0^T \frac{ dT}{T^{3}}\, e^{-m^2T}
e^{\I h A T}\,
\Big\langle e^{-g v[y] T} \Big\rangle_y,
\label{GQA5}
\end{equation}
where we have taken the photon field dependence out of the worldline
average, owing to the $A=$const. assumption.

We make use of the potential distribution, as introduced in the previous
section, since it is sufficient for
evaluating the nonperturbative effective action of the scalar photon
and performing the renormalization program. Let us illustrate this
statement with the aid of a simplified fit to the potential
distribution; a better but more complex fit will be presented
below. As a simplified parameterization of the potential distribution,
let us consider
\begin{equation}
P(v) = \frac{\beta^{1+\alpha}}{\Gamma(\alpha+1)} \, (v-v_0)^\alpha\,
 \exp(-\beta (v-v_0))\, \theta(v-v_0),
\label{eq:stos_pot_dis}
\end{equation}
where $\alpha,\beta$ and $v_0$ denote fit parameters and $\theta$ is
the step function. This distribution is already normalized, $\int dv\,
P(v)=1$. Numerically, we find that $\alpha$ and $\beta$ are finite,
$\alpha\simeq 0.79$ and $\beta\simeq 13.2$. Only $v_0$
diverges with the number of points per loop, which can be related to
the divergence of the potential ensemble average, since
\begin{equation}
<v>_y = \int dv \; v \; P(v)=v_0+ \frac{1+\alpha}{\beta}.\label{vvev}
\end{equation}
With regard to our result of Eq.~\eqref{vvev1}, this implies that
\begin{equation}
v_0=b \ln N + a-\frac{1+\alpha}{\beta}
\simeq 0.34 \, \ln N + 0.23 \,. \label{v_0}
\end{equation}
Most importantly, the distribution also provides us with a numerical
result for the exponential of the potential, carrying the
nonperturbative information about the infinite set of quenched
diagrams,
\begin{eqnarray}
\langle e^{-gT\, v[y]}\rangle_y &=& \int dv\, P(v)\, e^{-gT v}
\nonumber\\
&=& \left( \frac{\beta}{\beta+gT} \right)^{1+\alpha}\, e^{-gTv_0}
  \equiv F_{(\alpha\beta)}(gT)\, e^{-gTv_0}. \label{expvev}
\end{eqnarray}
For later convenience, we have introduced the auxiliary function
\begin{equation}
\quad
F_{(\alpha\beta)}(x)=\left(\frac{\beta}{\beta+x}\right)^{1+\alpha}.
\label{DefFa}
\end{equation}
Inserting this result into Eq.~\eqref{GQA5}, we obtain the numerical
estimate for the unrenormalized effective action,
\begin{equation}
\GQA[A]= -\frac{1}{32\pi^2} \int d^4x
\int_0^T \frac{ dT}{T^{3}}\,
e^{-m^2T} e^{\I h A T}\,
 F_{(\alpha\beta)}(gT)\, e^{-gTv_0}, \quad g=\frac{h^2}{8\pi^2}.
\label{GQA6}
\end{equation}
As for the scalar-photon operators, we demand the renormalization
conditions (independently of the approximations)
\begin{equation}
\frac{\delta \Gamma[A=0]}{\delta A(x)} =0, \quad
\frac{\delta^2 \Gamma[A=0]}{\delta A(-p)\delta A(p)}\bigg|_{p^2=0}
=0. \label{ARcond}
\end{equation}
The first condition implements the ``no-tadpole'' renormalization
prescription, whereas the second condition defines the scalar photon
to be massless. In general, a third condition fixing the wave function
renormalization of the scalar photon $Z_A$ has to be implemented; however,
it is a peculiarity of the present model that the renormalization
shift of this $Z_A$ factor remains zero $\delta Z_A=0$. \footnote{This
  is reminiscent of the same relation of $\phi^4$ theory in $D=4$ to
  one-loop order; because of the super-renormalizability of the present
  model, this relation holds to all orders here.} Incidentally, the
same is true for the wave function renormalization of the $\phi$
field.

These renormalization conditions fix the corresponding counterterms,
resulting in the following (partially renormalized) effective action:
\begin{equation}
\GQA[A]= -\frac{1}{32\pi^2} \int d^4x\, \int_0^T \frac{ dT}{T^{3}}\,
e^{-m^2T} \left(e^{\I h A T}-1-\I hAT+ \frac{(hAT)^2}{2} \right)\,
 F_{(\alpha\beta)}(gT)\, e^{-gTv_0},
\label{GQA7}
\end{equation}
where we have also subtracted the zero-point energy, enforcing
$\Gamma[A=0]=0$. Note that these renormalization conditions have
rendered the propertime integral finite. This is, in particular,
independent of the loop order in the quenched approximation, since the
counterterms do not receive contributions from higher-loop orders due
to super-renormalizability.\footnote{This point is expected to be
  different in renormalizable theories, such as QED.}.

The remaining divergence contained in $v_0$ can finally be removed by
mass renormalization. In principle, this is done by imposing a
renormalization condition for the 2-point function of the $\phi$ field
at zero momentum. This is straightforwardly possible within our approach
but requires an explicit computation of the 2-point function. However,
we adopt a simpler prescription here for which it suffices to consider
merely the scalar photon action: we demand that the propertime
integrand should fall off exponentially for large $T$ with a width
given by the renormalized mass,
\begin{equation}
\frac{1}{T^3}\, e^{-m^2T} \left(e^{\I h A T}-1-\I hAT+
 \frac{(hAT)^2}{2} \right)\,   F_{(\alpha\beta)}(gT)\,
 e^{-gTv_0}\bigg|_{T\to\infty}
\sim f(T)\, e^{-m_{\text{R}}^2 T}, \label{massren}
\end{equation}
where the prefactor $f(T)$ has a weaker $T$ dependence than an
exponential function. This renormalization condition also fixes the
finite part of the mass renormalization according to
\begin{equation}
m_{\text{R}}^2=m^2+\frac{h^2}{8\pi^2}\, v_0=m^2+ g \left( b \ln N + a
-\frac{1+\alpha}{\beta} \right). \label{massren2}
\end{equation}
We end up with the fully renormalized effective action for soft scalar
photons in the quenched approximation,
\begin{equation}
\GQAr[A]=-\frac{1}{32\pi^2} \int d^4 x
\int_0^T \frac{ dT}{T^{3}}\,e^{-m_{\text{R}}^2T}
\left(e^{\I h A T}-1-\I hAT+ \frac{(hAT)^2}{2} \right)\,
 F_{(\alpha\beta)}\left( \frac{h^2}{8\pi^2}T\right)\,,
\label{GQA8}
\end{equation}
where the form of the function $F_{(\alpha\beta)}(x)$ is given in 
Eq.~(\ref{DefFa}) and
holds for our simplified fit \eqref{eq:stos_pot_dis}.

\begin{figure}[hbt]
\centering
\mbox{\epsfig{figure=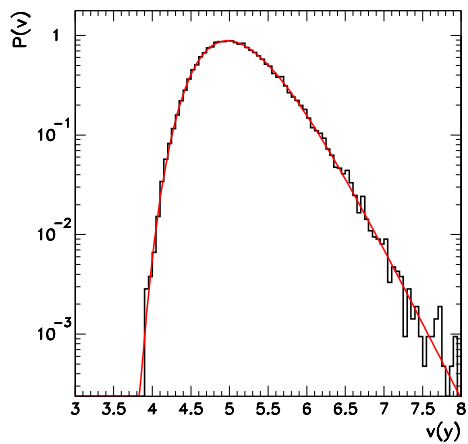,angle=0,width=7.0cm}}
\caption{Probability distribution of the self-interaction potential
  (histogram) and a fit to a modified log Gau{\ss}ian (continuous line).}
\label{fig:vstos}
\end{figure}

A more accurate parameterization of the potential distribution can be
obtained by the fit function
\begin{equation}
P(v) = p_1 \exp \{ -\frac{1}{2 p_2^2} (\log(v-v_0)-p_3)^2 \} \,
\theta(v-v_0),
\label{eq:stos_pot_acc}
\end{equation}
a plot of which is shown in Fig.~\ref{fig:vstos} together with the
potential distribution for $N= 2^{14}$.  The $p_i$ as well as $v_0$
are the fit parameters. Again, we observe that only $v_0$ carries a
logarithmic divergence, whereas the $p_i$ are finite numbers. From the
fit we obtain $p_2 \simeq 0.24$, $p_3 \simeq 0.60$, and
$p_1\simeq0.89$ fixes the overall normalization. $v_0$ is related to
the average potential 
\begin{equation}
v_0 = <v>_y - \frac{1}{\sqrt \pi} e^{p_3 + 3/2 p_2^2} \simeq
0.34 \ln(N) -0.76 \, .
\end{equation}

This improved fit results in a better numerical estimate of the
potential average,
\begin{eqnarray}
\langle e^{-gTv[y]}\rangle_y&=&e^{-gT v_0} \,
   p_1 \int_0^\infty dv\, e^{-gTv}\, e^{-\frac{1}{2p_2^2}
   (\ln(v)-p_3)^2} \nonumber\\
&=:& e^{-gT v_0}\, F_{(p_i)}(gT), \label{betterfit}
\end{eqnarray}
which is of the same form as for the simplified fit in Eq.~\eqref{expvev}, but
involves a more complex function $F_{(p_i)}(x)$ defined by a parameter
integral. The mass renormalization proceeds as above, $m_{\text{R}}^2 = m^2
+h^2/(8\pi^2)\,v_0$, since $F_{(p_i)}(gT)$ decays weaker than exponentially
for large $T$, see App.~\ref{AppFitF}. Therefore, replacing
$F_{(\alpha\beta)}(x)$ in Eq.~\eqref{GQA8} by $F_{(p_i)}(x)$, we obtain our
improved result for the effective action for the scalar photon.

At this point, it is important to stress that both fits do not only
differ with respect to their quality of describing the data, but also
with respect to their analytic properties. This becomes particularly
apparent from the corresponding perturbative expansions. For instance,
the simple fit leads to $F_{(\alpha\beta)}(gT)$, Eq.~\eqref{DefFa},
which is completely analytic in $g$ and hence can be expanded in a
well-converging Taylor series, as displayed in Eq.~\eqref{smallgTsim}
in App.~\ref{AppFitF}. By contrast, the resulting auxiliary function
$F_{(p_i)}(gT)$ from the better fit gives rise to an asymptotic series
upon expansion for $g\to 0^+$, as shown in Eq.~\eqref{smallgT}.

Within the language of diagrammatic perturbation theory, the
implications from the two fits are different. To see this, let us note
that the ratio of successive loop contributions to a scalar photon
proper vertex (obtained by expanding $\GQA[A]$ in terms of $A$) is
proportional to the ratio of the corresponding expansion coefficients
of the auxiliary function,
\begin{equation}
\frac{\GQAr^{(n)}|_{l+1\,\,\text{loop}}}
  {\GQAr^{(n)}|_{l\,\,\text{loop}}}\Bigg|_{\text{l.g. in}\,\,l}=
-(l+n)\, \frac{F^{(l)}(gT=0)}{F^{(l-1)}(gT=0)}\, g, \quad
    \GQAr^{(n)}=\frac{\delta^n
\GQAr[A=0]}{\delta A \dots \delta A}, \label{ratio}
\end{equation}
where $F^{(l)}(gT=0)$ denotes the $l$th expansion coefficient of the
corresponding auxiliary function (see, Eqs.~\eqref{smallgT} and
\eqref{smallgTsim}), and we have concentrated on the leading growth (l.g.)
of the ratio for large $l$. The prefactor of $(l+n)$ arises from the
propertime integration and can be traced back to the fact that the
self-interaction potential $V$ is dimensionful in the scalar
model.\footnote{This is different in ScQED where the potential is
dimensionless in $D=4$; hence, this factor $\sim l$ would be absent.}
Now, for the two different fits, this coefficient
ratio reads
\begin{equation}
\frac{\GQAr^{(n)}|_{l+1\,\,\text{loop}}}
  {\GQAr^{(n)}|_{l\,\,\text{loop}}}\Bigg|_{\text{l.g. in}\,\,l}=
-g \,\times \left\{
\begin{array}{ll}
 \frac{l+n}{\beta} & \text{for}\quad F_{(\alpha\beta)} \\
 e^{p_2^2(l+1/2)-p_3} (1+\frac{n}{l})& \text{for}\quad F_{(p_i)}
\end{array}\right.
. \label{ratioforfit}
\end{equation}
This implies that the proliferation of diagrams with increasing loop
order translates to an increase of the expansion coefficient which is
very different for the two fits. The simple fit, Eq.~\eqref{eq:stos_pot_dis}, 
leads to a mild
increase of the expansion coefficients. Though the resulting
perturbative series would not be convergent but asymptotic, the series
would be immediately Borel summable.\footnote{In the somewhat
artificial large-$n$ (large photon number) limit, i.e., $n\gg l$, the
series is indeed absolutely convergent for small coupling, being of
the type of a geometric series.} For instance, for the case of ScQED,
it has recently been conjectured that the perturbation series even
converges for the on-shell renormalized QED $n$-photon amplitudes in
the quenched approximation \cite{Dunne:2004xk}.

For the better fit, Eq.~\eqref{eq:stos_pot_acc}, the coefficient ratio 
increases strongly, implying
that a resummation requires generalized resummation techniques. Since
the leading-growth behavior of the coefficients from the loop
expansion are influenced by the combinatorics of diagrams, the
conclusion for this perturbative combinatorics would be very different
for the two fits. The lesson to be learned is that drawing a reliable
conclusion about the analyticity structure of the perturbative
expansion from the numerical estimate is very difficult. Different
fits of similar quality can lead to very different analytic
structures. However, since we have the fully resummed finite result
numerically at our disposal, an analysis of the analytic structure may
be viewed as a rather academic question.
Notice also that the large $gT$ expansion is controlled by the small $v$
behavior, {\it i.e.}, it is related to the approach to the ``classical''
region where $v\simeq v_{\rm min}$. This classical region corresponds
to a subset of worldlines that are close to the minima of the action
and thus resemble classical trajectories. 

Let us finally point out one remarkable conclusion that can be drawn
form the effective action \eqref{GQA8}. To one-loop order, the
small-mass or large-photon-field limit of the effective action is of
the form
\begin{equation}
\Gamma_{\text{1-loop}}[A]\big|_{\frac{hA}{m_{\text{R}}^2} \gg 1} 
\simeq- \frac{1}{64\pi^2}
\int d^4x\, (hA)^2 \ln \frac{hA}{m_{\text{R}}^2}, \label{oneloop}
\end{equation}
exhibiting a logarithmic increase, as is familiar from the
Heisenberg-Euler QED effective action or the Coleman-Weinberg
potential. This form implies that the massless $\phi$ field limit is
not well defined for our scalar model; technically speaking, the $T$
integral is divergent at the upper bound in the massless limit. This
becomes different, once the infinite number of diagrams of the
quenched approximation is taken into account. These induce the
$F_{(\dots)}(gT)$ function which vanishes for larger $T$,
$F_{(\dots)}(gT\to \infty)\to 0$. This property renders the $T$
integrand finite even in the massless $\phi$ field limit. For
instance, for the simple fit, the massless limit of $\GQA[A]$ can be
evaluated analytically. Here, we just cite the result for the limit
$A/h\ll 1$, for which the quenched approximation is expected to be
more reliable,
\begin{equation}
\GQAr[A]|_{m_{\text{R}}=0} =- \frac{[-\Gamma(-2-\alpha)] \cos
  \frac{\pi}{2} \alpha}{2^{5-3\alpha} \pi^{2(1-\alpha)} \beta^\alpha}
\int d^4x\, (hA)^2\, \left(\frac{A}{h}\right)^\alpha
 [1+ \mathcal{O} ((A/h)) ]  .
\label{beyoneloop}
\end{equation}
The overall sign is, of course, still negative since
$\alpha\simeq0.79$.  We conclude that a completely massless scalar
model of the present type can be formulated consistently in the
quenched approximation; it is only the perturbative expansion of this
model that breaks down in the massless limit.

\section{Conclusions and Outlook}
\label{SecConc}

The feasibility of a nonperturbative analysis in quantum field theory
based on the worldline formalism has been demonstrated here. The
approach combines the efficiency of worldline representations, i.e., having
infinitely many Feynman diagrams in one worldline expression, with
powerful numerical algorithms for the worldline integrals. As a major
advantage of this approach, all continuous spacetime symmetries remain
unaffected by numerical discretization procedures.

In the present exploratory study, we have considered a scalar model
with cubic interaction in four dimensions in the quenched
approximation which is the leading non-trivial order in a small-$\Nf$
expansion.  This approximation is expected to hold for small external
field amplitudes but arbitrary values of the coupling.  The model
mimics scalar QED or Yukawa theory and is super-renormalizable though
not finite. Therefore the divergence structure is rich enough to allow
for a detailed study of the main difficulty of many field theoretic
methods: understanding the origin and taming of divergencies. The
combination of the propertime representation and the worldline
discretization provides for a sufficient regularization, facilitating
the identification of divergencies. 
Subsequently, the divergencies can be removed by fixing
physical parameters by virtue of renormalization conditions on a
nonperturbative level. It is particularly such a nonperturbative
analysis of quantum field theories which sheds light on the
fundamental problem whether a perturbative formulation gives an
exactly correct account of divergencies \cite{GlimmJaffe}. For
practical purposes, our divergence analysis may be helpful for
technically similar problems, e.g., arising in many effective theories
for polymer physics \cite{David}. 

For the case of the effective action for soft scalar photons, we have
performed this program explicitly. The result is a finite propertime
parameter integral of Heisenberg-Euler type, containing all-order
corrections in the coupling. These all-order contributions
particularly modify the large-propertime behavior, in turn affecting
the small-mass limit. We observe that the zero-mass limit surprisingly
exists in the nonperturbative result, whereas perturbation theory
is ill-defined in this limit. Hence, we conclude that the large
logarithms for small masses from all quenched diagrams can be summed
up to a finite result. Whether this feature persists beyond the
quenched approximation remains an open issue. 

At this point, a more detailed discussion of the interplay between the
various parameters and approximation methods is in order. In the
scalar model, we have three massive scales, implying two dimensionless
parameters, e.g., $hA/m^2$ and $hA/h^2\equiv A/h$. In the limit
$A/h\gg 1$, radiative scalar photon exchanges are suppressed and the
perturbative loop expansion is expected to become reliable, almost
independently of the value of $hA/m^2$; of course, for $h A/m^2\gg1$,
infinitely many couplings to the external field have to be taken into
account. For $A/h\ll 1$, the systems is in the nonperturbative
regime. In the additional limit of $hA/m^2\ll 1$, higher $\phi$ loops
are suppressed by the mass threshold in comparison to radiative photon
exchanges; hence, the quenched approximation is reliable in this
limit. Whether this suppression of $\phi$ loops persists also in the
massless limit of the scalar model is an open question. The fact that
the massless limit exists in the quenched approximation may be taken
as an indication for an affirmative answer. At least, there is no
obvious reason for additional virtual $\phi$ loops to exhibit a
problematic infrared behavior; on the contrary, virtual momenta
flowing into the virtual loops rather serve as additional infrared
regulators. In this sense, the quenched approximation with one virtual
$\phi$ loop and soft zero-momentum external photons represents the
case of highest sensitivity to the massless limit. To summarize, we
expect the finiteness of the massless limit in the scalar model to
persist beyond the quenched approximation; whether further $\phi$
loops give quantitatively sizable contributions to this result has to
be checked by explicit computation. For this, the necessary inclusion
of additional internal matter loops into our formalism appears to be
feasible with our present technology. We would finally like to stress
that the massless limit of the scalar model may be a peculiarity
induced by the existence of a massive coupling, and it might not
translate to a similar property of renormalizable models such as QED
or Yukawa interactions. For instance in spinor QED, an interacting
massless limit of the soft-photon action is inhibited either by
triviality or by chiral symmetry breaking \cite{Gockeler:1997dn}.

From a technical perspective, we have developed further the
algorithmic tools of worldline numerics. Beyond those techniques that
so far have successfully been applied to perturbative computations,
the use of probability distributions renders the algorithms extremely
efficient. For instance, the effective action, i.e., the generating
functional of infinitely many soft photon amplitudes, has been derived
from one single distribution (of the
self-interaction potential). A further illustrative example for the
harmonic oscillator is given in App.~\ref{HO}. We are convinced that
the method will find a variety of applications, given the large number
of problems that are being tackled by propertime methods.

Our work paves the way for many generalizations: for instance, both
the worldline formalism as well as our worldline numerics are well
suited for a study of propagators in coordinate space, where the scale
set by the propagation distance represents one further technical
challenge. The study of propagators appears mandatory for an analysis
of renormalizable theories such as (scalar) QED in $D=4$. In these
cases, our simplified mass renormalization will no longer be
sufficient and the renormalized mass needs to be fixed with the aid of
the propagator's correlation length. 

Moreover, a generalization to gauge interactions will require a more
careful treatment as far as the discretization is concerned. Our
present method if applied to a gauge theory with a charged $\phi$
field would not satisfy current conservation which is essential for
maintaining gauge invariance. More refined algorithms are required
and are a subject of current development.

\section*{Acknowledgments} We thank O. Alvarez, M. Asorey, J.L. Cort\'es,
G.V.~Dunne, C. Fosco, J.M. Pawlowski and C.~Schubert for discussions.
J.S.-G. and R.A.V. thank MCyT (Spain) and FEDER (FPA2002-01161), and
Incentivos from Xunta de Galicia. We thank ``Centro de
Supercomputaci\'on de Galicia'' (CESGA) for computer support. H.G. is
grateful for the hospitality of the Departamento de F\'{\i}sica de
Part\'{\i}culas, Universidade de Santiago, and he acknowledges support by
the Deutsche Forschungsgemeinschaft (DFG) under contract Gi 328/1-3
(Emmy-Noether program).

\begin{appendix}

\section{Derivation of the effective action}
\label{AppQA}

Here, we derive the worldline representation of the Schwinger
functional and the effective action for the scalar photon field
$\Gamma[A]$, Eq.~\eqref{GQA1}. 
We start by citing the Schwinger functional as given in
Eq.~\eqref{SchwingerF} with the insertion of the worldline form of
$S_{\text{eff}}$ of Eq.~\eqref{Seff},
\begin{equation}
Z[J]\equiv e^{W[J]}=\int \mathcal D \hat A \, \exp \left[
  -\int_x \frac{1}{2} (\partial_\mu  \hat A)^2
  + \int_T \Big\langle e^{\int_x j \hat A} \Big \rangle_x
  +\int_x J \hat A\right],
\end{equation}
where we have used the worldline current representation of
Eq.~\eqref{jcurr}, and introduced the abbreviation
\begin{equation}
\int_T := \frac{1}{2(4\pi)^{D/2}} \int_0^T \frac{dT}{T^{1+D/2}}
           \,e^{-m^2 T}.
\label{intT}
\end{equation}
Let us first expand the worldline exponential, such that the $\hat A$
integral becomes Gau\ss ian at each order of the Taylor sum,
\begin{eqnarray}
Z[J]&=&\sum_0^\infty \frac{1}{n!} \int_{T_1} \dots \int_{T_n}
\left\langle \int \mathcal D \hat A\, e^{-\int_x \frac{1}{2} (\partial
  \hat A)^2} e^{\int_x \hat A (j_1+\dots+j_n+J)} \right\rangle_{x_1
  \dots x_n} \nonumber\\
&=&\sum_0^\infty \frac{1}{n!} \int_{T_1} \dots \int_{T_n}
\left\langle e^{\frac{1}{2} \int_x K_n^J \Delta K_n^J}
  \right\rangle_{x_1\dots x_n}\nonumber\\
&\equiv& e^{W[J]},\label{WJ}
\end{eqnarray}
where we have introduced the generating functional of connected
Green's functions $W[J]$ in the last line, and have defined the sum of
currents
\begin{equation}
K_n^J=J+\sum_i^n j_i. \label{KnJ}
\end{equation}
The kernel $\Delta$ denotes again the scalar photon propagator as
given in Eq.~\eqref{Delta}. Note that Eq.~\eqref{WJ} is an exact
representation of the Schwinger functional without any approximation
so far. The classical field $A$ (conjugate to the source $J$) is
obtained from $W[J]$ by functional differentiation
(cf. Eq.~\eqref{classA}),
\begin{eqnarray}
A&=& \frac{\delta W[J]}{\delta J}\nonumber\\
&=&\Delta\, J+e^{-W[J]} \sum_n \frac{1}{n!} \int_{T_1\dots T_n}
  \left\langle \Delta \sum_{i=1}^n j_i\,\, e^{\frac{1}{2} \int_x K_n^J
    \Delta K_n^J} \right \rangle_{x_1\dots x_n}. \label{AofJ}
\end{eqnarray}
A partial inversion of this equation gives,
\begin{equation}
J=\Delta^{-1}\, A - e^{-W} \sum_n \frac{1}{n!}\, \int_{T_1\dots T_n}
  \left\langle \sum_{i=1}^n j_i\,\, e^{\frac{1}{2} \int_x K_n^J
    \Delta K_n^J} \right \rangle_{x_1\dots x_n}, \label{JofA}
\end{equation}
with $\Delta^{-1}= -\partial^2$.  Note that this is an implicit
definition of $J=J[A]$, since $K_n^J$ in the exponent on the
right-hand side still depends on $J$. Nevertheless, these definitions
can be used to give an implicit definition of the exact effective
action for the scalar photons $\Gamma[A]$ (cf. Eq.~\eqref{Legendre}),
\begin{equation}
\Gamma[A]=-W[J[A]] + \int_x A\, J[A]. \label{Leg2}
\end{equation}
Explicit representations can be found in the quenched approximation
that consists of dropping all terms involving worldline correlators of
different worldline currents, $ \langle j_i \dots j_j\rangle_{x_i\dots
  x_j}\to 0$, $i\neq j$. In this approximation, the Schwinger
functional Eq.~\eqref{WJ} can be written as
\begin{eqnarray}
e^{\WQA[J]}&=&e^{\frac{1}{2} \int_x J\Delta J} \sum_n \frac{1}{n!}
\int_{T_1} \Big\langle e^{\frac{1}{2}\int_x j_1\Delta j_1 +\int_x
  J\Delta j_1} \Big\rangle_{x_1} \dots
\int_{T_n} \Big\langle e^{\frac{1}{2}\int_x j_n\Delta j_n +\int_x
  J\Delta j_n} \Big\rangle_{x_n} \nonumber\\
&=& \exp \left[ \frac{1}{2} \int_x J\Delta J +\int_T
  \Big\langle e^{\frac{1}{2}\int_x j\Delta j +\int_x
  J\Delta j} \Big\rangle_{x} \right]. \label{WJQA}
\end{eqnarray}
The corresponding representation of the classical field in quenched
approximation boils down to
\begin{equation}
A=\Delta\, J + \int_T \Big\langle e^{\frac{1}{2}\int_x j\Delta j +\int_x
  J\Delta j} \, \Delta\, j\Big\rangle_{x}. \label{classA2}
\end{equation}
Note that, even in the quenched approximation, this equation cannot
be inverted explicitly for the current,
\begin{equation}
J=\Delta^{-1} A -\int_T\Big\langle e^{\frac{1}{2}\int_x j\Delta j +\int_x
  J\Delta j} \, j\Big\rangle_{x}. \label{JofA2}
\end{equation}
We take this equation as the implicit definition of $J=J[A]$.  By
virtue of Eq.~\eqref{Leg2}, this allows to express the effective
action $\GQA[A]$ intermediately in terms of $J[A]$,
\begin{equation}
\GQA[A]= \frac{1}{2} \int_x J\Delta J + \int_T
  \Big\langle e^{\frac{1}{2}\int_x j\Delta j +\int_x
  J\Delta j} \,(J\Delta j -1 )\Big\rangle_{x}. \label{GQA3}
\end{equation}
Now, $J[A]$ of Eq.~\eqref{JofA2} can be iteratively inserted into
Eq.~\eqref{GQA3}. As a consequence, the iteration terminates after
the first order by means of the quenched approximation, since the
higher orders involve multi-worldline current correlators. The result
can be displayed in closed form,
\begin{equation}
\GQA[A]=\int_x \frac{1}{2} (\partial_\mu A)^2 -\frac{1}{2(4\pi)^{D/2}}
\int_0^\infty \frac{dT}{T^{1+D/2}} \Big\langle
  e^{\frac{1}{2}\int_x j\Delta j +\int_x
   j A} \Big\rangle_{x}, \label{derGQA}
\end{equation}
which serves as the starting point of our investigation in
Eq.~\eqref{GQA1}.

\section{Worldline generation: ``d loops''}
\label{Appdloops}

Here we describe a new algorithm for generating open or closed
worldlines {\em ab initio}, which is used for most numerical studies
in the present work. The algorithm generates worldlines by a suitable
doubling of points, starting from a small number (e.g., from 1
point); hence we call the resulting worldlines ``d loops'' (closed) or
``d lines'' (open). The algorithm is reminiscent of a heat-bath
algorithm, as was used in earlier worldline numerical studies
\cite{Gies:2001zp}, but does not require dummy thermalization sweeps.

Let us first consider closed ``d loops''. By appropriate rescaling,
the Gau\ss ian velocity distribution can always be brought to {\em
  unit loop} form,
\begin{equation}
P[y]=e^{-\frac{1}{4} \int_0^1\, dt\,
  \left(\frac{dy(t)}{dt}\right)^2}. \label{unitloops}
\end{equation}
Discretizing the $t$ derivative by an asymmetric nearest-neighbor
form, $dy/dt=(y_i-y_{i-1})/\Delta t$, with $i=1, \dots, N_0$, the
distribution becomes
\begin{equation}
P_{N_0}[y]\to e^{-\frac{N_0}{4} \sum_{i=1}^{N_0}\,
  (y_i-y_{i-1})^2}. \label{discdist}
\end{equation}
The probability for the location of the $i$'th point is
\begin{equation}
p_{y_i}=e^{-\frac{N_0}{4} [2 y_i^2-2y_i(y_{i+1}+y_{i-1})] \dots}
=e^{-\frac{N_0}{4}\, 2 [ y_i-\frac{1}{2}(y_{i+1}+y_{i-1})]^2 \dots}
,\label{yidist}
\end{equation}
such that $y_i$ is in a Gau\ss ian heat bath of the algebraic mean of
its nearest neighbors. The crucial point now is that we can add
another $N_0$ points $y_{i'}$ in between with the following probability
distribution,
\begin{equation}
p_{y_{i'},y_{i'+1}}=e^{-\frac{2N_0}{4} [2(y_{i'}
    -\frac{1}{2}(y_i+y_{i-1}))^2 +2(y_{i'+1}
    -\frac{1}{2}(y_{i+1}+y_{i}))^2 ]\dots},  \label{inbetween}
\end{equation}
such that the new total probability distribution for $y_i$ becomes
\begin{equation}
p_{y_i}\to p_{y_i} p_{y_{i'},y_{i'+1}}
=e^{-\frac{2N_0}{4}\, 2\, [y_i-\frac{1}{2}(y_{i'+1}+y_{i'})]^2
  \dots}.
\label{pyinew}
\end{equation}
Note that the probability distribution for the doubled points $y_{i'}$
in Eq.~\eqref{inbetween} has been adjusted such that the probability
distribution for the point $y_i$ has lost its dependence on $y_{i+1}$
and $y_{i-1}$; $y_i$ only feels the heat bath of the new points
$y_{i'}$ and $y_{i'+1}$. Correspondingly, the full distribution for
the $N_0+N_0=2N_0=N_1$ points reads:
\begin{equation}
P_{N_1}[y]\to e^{-\frac{N_1}{4} \sum_{i=1}^{N_1}\,
  (y_i-y_{i-1})^2}, \label{discdist2}
\end{equation}
where we have redefined $y_i\to y_{2i}$, $y_{i'}\to y_{2i-1}$.

Now the recipe for the ``d loop'' algorithm generating closed
worldlines with $N$ ppl is obvious:

\begin{itemize}
\item[(1)] Begin with one arbitrary point $N_0=1$, $y_{N}$.

\item[(2)] Create an $N_1=2$ loop, i.e., add a point $y_{N/2}$ that is
  distributed in the heat bath of $y_N$ with
\begin{equation}
e^{-\frac{N_1}{4} 2 (y_{N/2} -y_{N})^2}. \label{yn2}
\end{equation}

\item[(3)] Iterate this procedure, creating an $N_k=2^k$ppl loop by
  adding $2^{k-1}$ points $y_{{qN}/{N_k}}$, $q=1,3,\dots, N_k-1$ with
  distribution
\begin{equation}
e^{-\frac{N_k}{4} 2 [y_{qN/N_k} -\frac{1}{2}(y_{(q+1)N/N_k}+
    y_{(q-1)N/N_k})]^2}. \label{ynk}
\end{equation}

\item[(4)] Terminate the procedure if $N_k$ has reached $N_k=N$ for
  unit loops with $N$ ppl.

\item[(5)] For an ensemble with common center of mass, shift each
  whole loop accordingly.

\end{itemize}

Open worldlines (``d lines'') can be generated analogously by starting
with the end points $\xI=y_0$ and $\xF=y_{N+1}$ and filling $N$ new
points in between by the same doubling procedure.

In comparison with existing worldline generator algorithms such as the
``f loop'' and ``v loop'' algorithms of \cite{Gies:2003cv}, all algorithms
share the property that each generated random number is used for a
worldline; in particular, no dummy thermalization sweeps are
necessary. Similarly to the ``f loop'' algorithm, ``d loops'' produce
worldlines with $N=2^{l}$ ppl, $l\in \mathbbm{N}$, whereas the ``v
loop'' algorithm can produce any $N\in \mathbbm{N}$. ``f loops'' by
Fourier transformation discretize the derivative $\dot x$ implicitly
such that the operator $d^2/d\tau^2$ is diagonal in Fourier space,
whereas ``v loops'' and ``d loops'' use the asymmetric
nearest-neighbor discretization $\dot x\sim x_i-x_{i-1}$. The later
has proved to approach the continuum limit in problems with scalar
interactions faster than the Fourier decomposition \cite{Gies:2003cv}. In
comparison to ``v loops'', the present ``d loop'' algorithm requires
less algebraic operations and hence is faster by some ten percent.

\section{Harmonic oscillator}
\label{HO}

\begin{figure}[hbt]
\centering
\mbox{\epsfig{figure=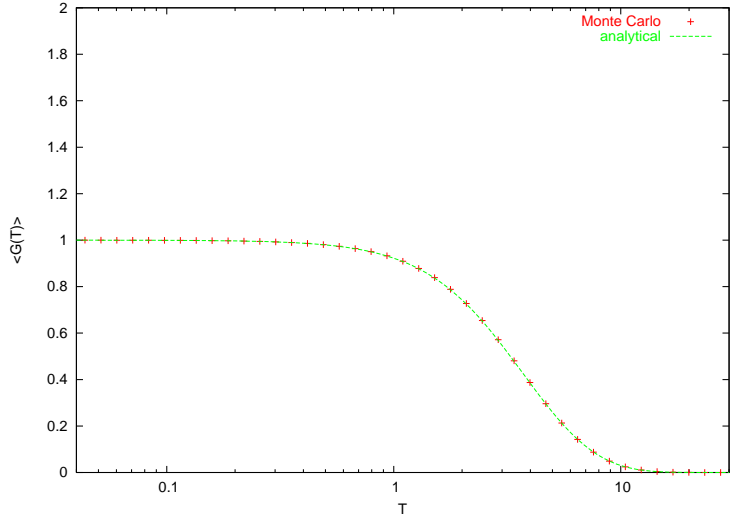,width=7.0cm,angle=-90}}
\caption{Propagator for the harmonic oscillator for $\xI=\xF=0$,
for $N=2^{14}$ ppl and $\omega=1$ and 10$^4$ worldline ensembles.
The Monte-Carlo calculation (points) is shown in comparison to
the analytical calculation of Eq.~\eqref{HOana} (line). }
\label{fig:armonic}
\end{figure}

As an instructive example of the use of the Monte-Carlo method involving
distribution fits, let us study the harmonic oscillator. The propagator for
a harmonic oscillator can be written in the form
\begin{equation}
G(T,R) = \int_{x(0)=\xI}^{x(T)=\xF} {\cal D}x \exp \{ -\frac{1}{2} \; \int_0^T
d\tau \; \dot x^2 + \omega^2 x^2 \}, \quad R=|\xF-\xI|.
\end{equation}
In the language of quantum field theory, this corresponds to the heat kernel
of a $D=1$ dimensional scalar field theory, coupling to a constant
``potential'' $\omega^2$. In the notation of worldline ensemble averages, this
propagator can be written as\footnote{The normalization of the kinetic term
  here is different from the main text, in order to make contact with the
  standard quantum mechanical normalizations. Both normalizations are
  connected by a simple rescaling of the worldlines.}
\begin{equation}
G(T,R) = \left\langle \exp \left[ - \frac{1}{2} \int_0^T d\tau\, \omega^2 x^2
  \right] \right\rangle_{\xI}^{\xF}.
\end{equation}
Introducing the $T$ scaling as outlined in Eq.(\ref{unitl1}) yields
\begin{equation}
G(T,R) = \Big\langle \exp \left[ - \frac{1}{2} \omega^2 T^2 \int_0^1 dt \;
  y^2 \right] \Big\rangle_{\sqrt{T}y_{\text{I}}}^{\sqrt{T}y_{\text{F}}}.
\end{equation}
The integral is Gau{\ss}ian and can be done analytically  \cite{Itzykson}. For
$\xI=\xF=0$, it is simply given by
\begin{equation}
G(T,R=0) = \sqrt \frac{\omega T}{\sinh \omega T}. \label{HOana}
\end{equation}
In Fig.~\ref{fig:armonic}, we plot the result of our Monte-Carlo
calculation with $N = 2^{14}$ ppl for $\omega=1$.

\begin{figure}[hbt]
\centering
\mbox{\epsfig{figure=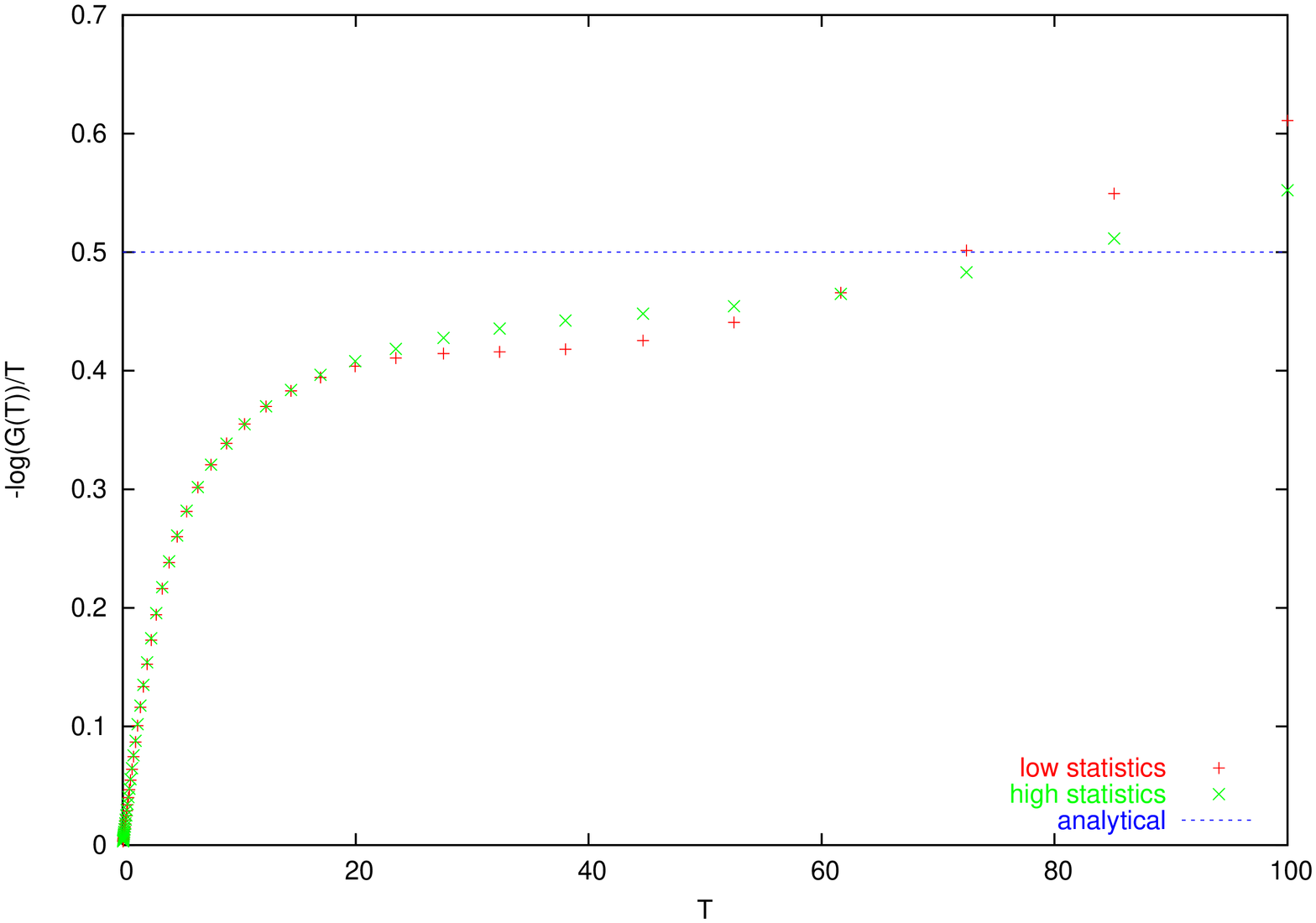,width=7.0cm,angle=-90}}
\caption{$-\log G(T) /T$ for the harmonic oscillator. Shown is the
  exact analytical result for $\omega=1$ (dashed line) and the Monte-Carlo
  results with $2^{10}$ ppl for $10^4$ and $10^6$ worldline ensembles.}
\label{fig:ground}
\end{figure}

The agreement with the analytical result is excellent, coinciding for more
than 10 orders of magnitude with a very small error.
However, the Monte-Carlo calculation is not free from problems, as becomes
visible, e.g., for the ground state energy. The latter can be calculated from
the standard identity
\begin{equation}
E_0 = \lim_{T \rightarrow \infty} -\frac{\log(G(T))}{T}.
\end{equation}
In Fig.~\ref{fig:ground}, we plot $-\log(G(T))/T$ versus $T$ for large
values of $T$, comparing the analytical result to the Monte-Carlo
calculation with $\nL=10^4$ and $10^6$ worldline ensembles.
As is obvious from the figure, the Monte-Carlo approaches the analytical
result for intermediate $T$, but ultimately departs from the analytical result
from some larger value of $T$ on. This point depends on the statistics of the
Monte-Carlo. This phenomena is well known in the literature and is usually
referred to as the ``classical collapse'' \cite{Rafelski}.

We can cast the problem into our language by introducing the
distribution of potentials, i.e.,
\begin{equation}
G(T,R=0) = \int dv_2 P(v_2) \exp \left[ - \frac{1}{2} \omega^2 T^2
  v_2\right], \label{eq:armonicpot}
\end{equation}
where the self-interaction potential is a local expression here:
\begin{equation}
v_2 = \int_0^1 dt \;  y^2.
\end{equation}
\begin{figure}[hbt]
\begin{center}
\epsfig{figure=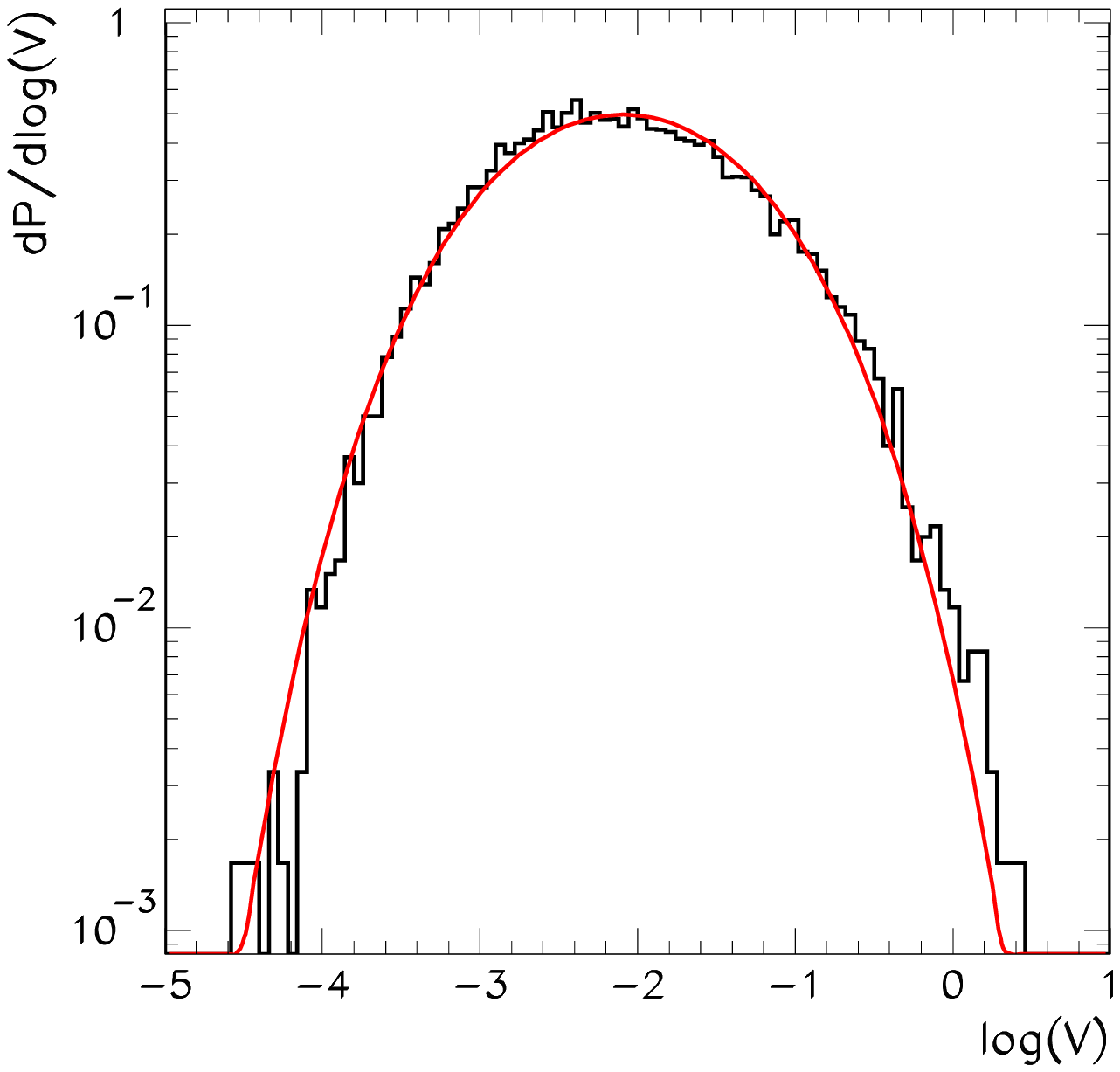,width=7.0cm}
\end{center}
\caption{Potential distribution $P(v_2)$ for the harmonic oscillator for
  $N=2^{10}$ ppl and 10$^6$ worldline ensembles. Also shown is the fit
  to the proposed distribution.}
\label{fig:dual}
\end{figure}
From Eq.~\eqref{eq:armonicpot}, we observe that the limit $T \rightarrow
\infty$ is controlled by $v_2 \rightarrow 0$. However, the distribution of
potentials $P(v_2)$ in this small $v_2$ limit is poorly determined, owing to
the lack of statistics, $P(v_2) \rightarrow 0$. A typical distribution
$P(v_2)$ is displayed in Fig.~\ref{fig:dual} for $N=2^{14}$ ppl. For any fixed
number of worldline ensembles, there is always a minimum value of the
potential, $v_{2\,{\rm min}}$, such that for any $T$ larger than that
determined by
\begin{equation}
\frac{1}{2} \omega^2 T^2 v_{2{\rm min}} \sim 1,
\end{equation}
the potential distribution $P(v_2)$ is not well represented by the given
statistical ensemble. This can be observed in Fig.~\ref{fig:ground}, where an
increase of the size of the ensemble enlarges the region in $T$ for which the
Monte-Carlo calculation is valid.

Our alternative method circumvents this problem by using an analytical
parameterization of $P(v_2)$ which has an appropriate behavior for low $v_2$.
For instance, the normalized trial function
\begin{equation}
P(v_2) = \hat{\mathcal N}\, e^{ -( a/v_2 + bv_2) }, \quad \hat{\mathcal
  N}^{-1}= 2\sqrt{\frac{a}{b}} \, K_1(2\sqrt{ab}),
\end{equation}
gives a reasonably good description of
$P(v_2)$ and reproduces the large $T$ behavior. Here $a \simeq 0.11$
and $b \simeq 18.4$ are constants determined by a fit to the potential
distribution.

From Eq.~\eqref{eq:armonicpot}, we immediately get a numerical estimate for
the propagator,
\begin{equation}
G(T,R=0)\big|_{P(v_2)} =\sqrt{\frac{2 b}{2b+\omega^2 T^2}} \,
\frac{K_1(2\sqrt{ab+a \omega^2 T^2/2})}{K_1(2\sqrt{ab})}, \label{propest}
\end{equation}
such that the large $T$ limit gives the ground state energy,
\begin{equation}
E_0\big|_{P(v_2)}=\sqrt{2 a}\, \omega, \label{E0est}
\end{equation}
which is in good agreement with the exact answer.

\section{Analysis of the potential fit function}
\label{AppFitF}

Here we analyze some properties of the auxiliary function
$F_{(p_i)}(x)$ introduced in Eq.~\eqref{betterfit} which are relevant
for the mass renormalization as well as the weak- and strong-coupling
behavior of the effective action for the scalar photon in
Sect.~\ref{SecGQA}.  This function is defined by 
\begin{equation}
F_{(p_i)}(x)=p_1 \int_0^\infty dv\, e^{-xv}\, e^{-\frac{1}{2p_2^2}
   (\ln(v)-p_3)^2},\label{DefF}
\end{equation}
with the $p_i$ being constant fit parameters and $x\equiv g T$ as in
the main text. We observe that the exponent becomes large (but
negative) for small as well as large $v$ with an extremum in
between. This suggest to evaluate the integral approximately by
steepest descent in order to obtain analytic information about its
behavior, e.g., for larger propertimes $T$. To be precise, we perform
the approximation to the following order,
\begin{eqnarray}
F_{(p_i)}(x)&=&p_1 \int_0^\infty dv\, e^{-f(v)} \nonumber\\
&\simeq& p_1 e^{-f(\vE)} \int_0^\infty dv\ e^{-\frac{1}{2} (v-\vE)^2\,
  f''(\vE)+ \dots}, \label{SPA}
\end{eqnarray}
where
\begin{equation}
f(v)=x\, v+ \frac{1}{2 p_2^2} [ \ln(v)-p_3]^2, \label{fofv}
\end{equation}
such that the extremum $\vE$ which is a minimum of $f(v)$ is
determined by $0=f'(\vE)$; the latter leads to a transcendental
equation
\begin{equation}
x \vE=\frac{p_3 -\ln(\vE)}{p_2^2}, \label{defve}
\end{equation}
that can be brought to a standard form,
\begin{equation}
p_2^2\, e^{p_3}\, x = w\, e^w, \quad \text{with}\quad
w=p_3-\ln \vE. \label{defve2}
\end{equation}
This form, $z=w\, e^w$, is the defining equation of the ``product
logarithm'' $w(z)$ by which we can express the extremum of the
integrand as a function of $x$:
\begin{equation}
\vE(x) = \frac{1}{p_2^2}\, \frac{w(p_2^2 e^{p_3}\,
  x)}{x}. \label{defve3}
\end{equation}
One important property of the product logarithm is its asymptotic
behavior for large argument, $w(z)\to \ln z$ for $z\to\infty$, such
that the extremum goes to zero for large propertimes $x=gT$. By
contrast, the ``inverse width'' of the fluctuations around the
extremum diverges for large propertimes and vanishing extremum,
\begin{equation}
f''(\vE)= \frac{1}{p_2^2} \, \frac{1+p_3 -\ln \vE}{\vE^2},
\label{invwid}
\end{equation}
such that the steepest-descent approximation is expected to be
accurate in this limit. Putting it all together, the steepest-descent
approximation of Eq.~\eqref{DefF} becomes upon insertion of
Eqs.~\eqref{fofv}, \eqref{defve3} and \eqref{invwid} into
Eq.~\eqref{SPA},
\begin{eqnarray}
F_{(p_i)}(gT\gg 1)&\simeq&\sqrt{\frac{\pi}{2}} \frac{p_1}{p_2} (p_2^2
e^{p_3})^{-1/p_2^2} \, \frac{\ln^{1/2} gT }{(gT)^{1+(1/p_2^2)}} \, \nonumber\\
&&\times \exp\left\{-\frac{1}{p_2^2} \left[\ln \left( \frac{1}{p_2^2}
  \frac{\ln(p_2^2 e^{p_3} gT)}{gT} \right) -p_3 \right]^2
\right\}. \label{SPA2}
\end{eqnarray}
Here we have already inserted the asymptotic form of the product
logarithm for large argument. In this large-$gT$ limit, the term $\sim
\exp (-(1/p_2^2) \ln^2 gT)$ dominats the decay of
$F_{(p_i)}(gT)$, which is weaker than exponential but stronger than
power-like. As a consequence, $F_{(p_i)}(gT)$ does not contribute any
finite part to the mass renormalization, due to the prescription
specified in Eq.~\eqref{massren}, but guarantees the finiteness of the
effective action $\GQA[A]$ in the massless-$\phi$ limit.

Let us now study the opposite limit of small $x=gT$ corresponding to a
perturbative expansion. For this, the first exponential in the
definition of $F_{(p_i)}(gT)$ in Eq.~\eqref{DefF} can be expanded and
the $v$ integral can be done order by order to give
\begin{equation}
F_{(p_i)}(gT)=\sqrt{2\pi}\, p_1 p_2\,\sum_{m=0}^\infty\,
\frac{e^{\frac{1}{2}(m+1)[(m+1)p_2^2+2p_3]}}{m!}\,
  (-gT)^m. \label{smallgT}
\end{equation}
The $m$th order in this expansion corresponds to the $l=m+1$-loop
order in perturbation theory.  For large $m$, the leading growth of
the expansion coefficient is $\sim \exp(m^2/2-m\ln m)$ by use of
Stirling's formula. Hence, this perturbative expansion gives an
asymptotic series in the coupling.

For comparison, we mention the analogous expansion for the simple-fit
result,
\begin{equation}
F_{(\alpha\beta)}(gT)=\sum_{m=0}^{\infty} \frac{1}{\beta^m}\,
\frac{\Gamma(m+\alpha+1)}{\Gamma(m+1)\Gamma(\alpha+1)}\, (-gT)^m,
\label{smallgTsim}
\end{equation}
implying a weaker leading growth of the expansion coefficient $\sim
m^\alpha$. This series is absolutely convergent for $|gT/\beta|<1$.

\end{appendix}


\begin{thebibliography}{99}
\setlength{\itemsep}{-0.3mm} 
{\small
\parskip=0pt

\bibitem{Heisenberg:1935qt}
W.~Heisenberg and H.~Euler,
Z.\ Phys.\  {\bf 98}, 714 (1936);\\
%
V. Weisskopf, K. Dan. Vidensk. Selsk. Mat. Fys. Medd. {\bf 14}, 1
(1936).

\bibitem{Coleman:1973jx}
S.~R.~Coleman and E.~Weinberg,
Phys.\ Rev.\ D {\bf 7}, 1888 (1973).

\bibitem{feynman1} R.P. Feynman, Phys. Rev. {\bf 80}, 440 (1950); {\bf
    84}, 108 (1951).

\bibitem{Polyakov:ez}
M.~B.~Halpern and W.~Siegel,
Phys.\ Rev.\ D {\bf 16}, 2486 (1977);\\
%
M.~B.~Halpern, A.~Jevicki and P.~Senjanovic,
Phys.\ Rev.\ D {\bf 16}, 2476 (1977);\\
%
%
A.~M.~Polyakov,
``Gauge Fields And Strings,'' Harwood, Chur (1987). 
%
\bibitem{berkos}
Z. Bern and D.A. Kosower, Nucl.~Phys.~{\bf B362}, 389
  (1991); {\bf B379}, 451 (1992).
%
\bibitem{strassler}
M.J.  Strassler, Nucl. Phys. {\bf B385},
  145 (1992).
%
\bibitem{Schmidt:1993rk}
M.~G.~Schmidt and C.~Schubert,
Phys.\ Lett.\ {\bf B318}, 438 (1993)
[hep-th/9309055];\\
M.~Reuter, M.~G.~Schmidt and C.~Schubert,
Annals Phys.\  {\bf 259}, 313 (1997)
[arXiv:hep-th/9610191];\\
R.~Shaisultanov,
Phys.\ Lett.\ B {\bf 378}, 354 (1996)
[arXiv:hep-th/9512142].
\bibitem{Schubert:2001he}
For a review, see C.~Schubert,
Phys.\ Rept.\  {\bf 355}, 73 (2001)
[arXiv:hep-th/0101036].

\bibitem{bedush}
Z. Bern, D.C. Dunbar, T. Shimada, Phys. Lett. {\bf
B 312} (1993) 277, hep-th/9307001.
\bibitem{dunnor}
D.C. Dunbar, P.S. Norridge,
Nucl. Phys. {\bf B 433} (1995) 181, hep-th/9408014.
\bibitem{cadhdu}
D. Cangemi, E. D'Hoker, G. Dunne,
Phys. Rev. {\bf D 51} (1995) 2513, hep-th/9409113.
\bibitem{Dilkes:1995cu}
F.~A.~Dilkes and D.~G.~C.~McKeon,
Phys.\ Rev.\ D {\bf 53}, 4388 (1996)
[arXiv:hep-th/9509005].
\bibitem{adlsch}
S.L. Adler, C. Schubert, Phys. Rev. Lett. {\bf 77} (1996) 1695,
hep-th/9605035.
\bibitem{gussho}
V.P. Gusynin, I.A. Shovkovy,
Can. J. Phys. {\bf 74} (1996) 282, hep-ph/9509383;
J. Math. Phys. {\bf 40} (1999) 5406, hep-th/9804143.
\bibitem{Fosco:2003rr}
C.~D.~Fosco, J.~S\'anchez-Guill\'en and R.~A.~V\'azquez,
Phys.\ Rev.\ D {\bf 69}, 105022 (2004)
[arXiv:hep-th/0310191].
\bibitem{Bastianelli:2004zp}
F.~Bastianelli and C.~Schubert,
JHEP {\bf 0502}, 069 (2005)
[arXiv:gr-qc/0412095].
\bibitem{marusc}
L. Magnea, R. Russo, and S. Sciuto, CERN-PH-TH/2004-244,
hep-th/0412087.
\bibitem{Brummer:2004xc}
F.~Brummer, M.~G.~Schmidt and Z.~Tavartkiladze,
arXiv:hep-th/0412284.

\bibitem{Gies:2001zp}
H.~Gies and K.~Langfeld,
Nucl.\ Phys.\ B {\bf 613}, 353 (2001)
[arXiv:hep-ph/0102185]; 
%
Int.\ J.\ Mod.\ Phys.\ A {\bf 17}, 966 (2002)
[arXiv:hep-ph/0112198].

\bibitem{Schmidt:2002mt}
M.~G.~Schmidt and I.~O.~Stamatescu,
Nucl.\ Phys.\ Proc.\ Suppl.\  {\bf 119}, 1030 (2003)
[arXiv:hep-lat/0209120]; 
%
Mod.\ Phys.\ Lett.\ A {\bf 18}, 1499 (2003).

\bibitem{Langfeld:2002vy}
K.~Langfeld, L.~Moyaerts and H.~Gies,
Nucl.\ Phys.\ B {\bf 646}, 158 (2002)
[arXiv:hep-th/0205304].

\bibitem{Gies:2003cv}
H.~Gies, K.~Langfeld and L.~Moyaerts,
JHEP {\bf 0306}, 018 (2003)
[arXiv:hep-th/0303264]; 
arXiv:hep-th/0311168.

\bibitem{WickCutkosky}
G.C.~Wick, Phys.\ Rev. {\bf 96}, 1124 (1954); \\%
R.E.~Cutkosky, Phys.\ Rev. {\bf 96}, 1135 (1954).

\bibitem{Affleck:1981bm}
I.~K.~Affleck, O.~Alvarez and N.~S.~Manton,
Nucl.\ Phys.\ B {\bf 197}, 509 (1982).

\bibitem{Rosenfelder:1995bd}
R.~Rosenfelder and A.~W.~Schreiber,
Phys.\ Rev.\ D {\bf 53}, 3337 (1996)
[arXiv:nucl-th/9504002];\\
C.~Alexandrou, R.~Rosenfelder and A.~W.~Schreiber,
Phys.\ Rev.\ A {\bf 59}, 1762 (1999)
[arXiv:hep-th/9809101]; 
C.~Alexandrou, R.~Rosenfelder and A.~W.~Schreiber,
Phys.\ Rev.\ D {\bf 62}, 085009 (2000)
[arXiv:hep-th/0003253].


\bibitem{Sanchez-Guillen:2002rz}
J.~S\'anchez-Guill\'en and R.~A.~V\'azquez,
Phys.\ Rev.\ D {\bf 65}, 105001 (2002)
[arXiv:hep-th/0201065].

\bibitem{Savkli:1999rw}
C.~Savkli, J.~Tjon and F.~Gross,
Phys.\ Rev.\ C {\bf 60}, 055210 (1999)
[Erratum-ibid.\ C {\bf 61}, 069901 (2000)]
[arXiv:hep-ph/9906211]; 
arXiv:nucl-th/0404068.

\bibitem{Brambilla:1997ky}
N.~Brambilla and A.~Vairo,
Phys.\ Rev.\ D {\bf 56}, 1445 (1997)
[arXiv:hep-ph/9703378].
%
\bibitem{Alvarez:1983}
L. Alvarez-Gaume, Commum. Math. Phys. {\bf 90}, 161, (1983).
%

\bibitem{Fosco:2004} 
C. Fosco, [arXiv:hep-th/0404068].
%

\bibitem{Bender:1999ek}
C.~M.~Bender, K.~A.~Milton and V.~M.~Savage,
Phys.\ Rev.\ D {\bf 62}, 085001 (2000)
[arXiv:hep-th/9907045].


\bibitem{GiesHaemmerling}
H.~Gies and J.~H\"ammerling,
[arXiv:hep-th/0505072].

%
\bibitem{Fosco:prep}
C. Fosco, J.S. S\'anchez-Guill\'en, and R.A. V\'azquez, in preparation
(2005). 
%

\bibitem{Dunne:2004xk}
G.~V.~Dunne and C.~Schubert,
[arXiv:hep-th/0409021].

\bibitem{Gockeler:1997dn}
M.~Gockeler, R.~Horsley, V.~Linke, P.~Rakow, G.~Schierholz and H.~Stuben,
Phys.\ Rev.\ Lett.\  {\bf 80}, 4119 (1998)
[arXiv:hep-th/9712244];\\
H.~Gies and J.~Jaeckel,
Phys.\ Rev.\ Lett.\  {\bf 93}, 110405 (2004)
[arXiv:hep-ph/0405183].


\bibitem{Itzykson} C. Itzykson and J.B. Zuber, {\it Quantum field
  theory},  Mc Graw Hill, 1980, New York. 
\bibitem{Rafelski} S.K. Kauffmann and J. Rafelski, Z. Phys. C {\bf 24},
 157 (1984).
%
\bibitem{GlimmJaffe}
J. Glimm and A.M. Jaffe, {\it Quantum Physics. A functional integral
  point of view}, New York, Springer (1981).
%
\bibitem{David}
F. David, K. Wiese, [arXiv:cond-mat/0409765].
}
\end{thebibliography}
\end{document}